\documentclass[11pt]{article}
\usepackage{multicol}
\usepackage{amssymb}
\usepackage{graphicx}
\usepackage{amsmath,amssymb}
\usepackage{amsthm}
\usepackage{array}
\usepackage{multirow}
\usepackage{booktabs,graphicx}

\newcommand{\mbf}[1]{{\mbox{\boldmath${#1}$}}}
\newtheorem{theorem}{Theorem}[section]

\newtheorem{corollary}[theorem]{Corollary}

\newtheorem{remark}{Remark}[section]

\newtheorem{assumption}{Assumption}

\begin{document}
\title{{\LARGE \textbf{A New Wald Test for Hypothesis Testing Based on MCMC outputs}}\thanks{%
Li gratefully acknowledges the financial support of the Chinese Natural
Science Fund (No. 71271221)¡£ Yu would like to acknowledge the
financial support from Singapore Ministry of Education Academic Research
Fund Tier 2 under the grant number MOE2011-T2-2-096 and Tier 3 under the
grant number MOE2013-T3-1-017. Yong Li, Hanqing Advanced Institute of
Economics and Finance, Renmin University of China, Beijing, 100872, P.R.
China. Email: gibbsli@ruc.edu.cn. Xiao-Bin, Liu, School of Economics,
Singapore Management University, 90 Stamford Road, Singapore 178903. Jun Yu,
School of Economics and Lee Kong Chian School of Business, Singapore
Management University, 90 Stamford Road, Singapore 178903. Email:
yujun@smu.ed.sg. URL: http://www.mysmu.edu/faculty/yujun/. }}
\author{Yong Li \\
\emph{Renmin University} \and Xiaobin Liu \\
\emph{Singapore Management University} \and Jun Yu \\
\emph{Singapore Management University} \and Tao Zeng\\
\emph{Wuhan University}}
\maketitle

\begin{abstract}
In this paper, a new and convenient $\chi^2$ wald test based on MCMC outputs is proposed for
hypothesis testing. The new statistic can be explained as MCMC version
of Wald test and has several important advantages that make it very
convenient in practical applications. First, it is well-defined under
improper prior distributions and avoids Jeffrey-Lindley's paradox. Second,
it's asymptotic distribution can be proved to follow the $\chi^2$
distribution so that the threshold values can be easily calibrated from this
distribution. Third, it's statistical error can be derived using the Markov
chain Monte Carlo (MCMC) approach. Fourth, most importantly, it is only
based on the posterior MCMC random samples drawn from the posterior
distribution. Hence, it is only the by-product of the posterior outputs and
very easy to compute. In addition, when the prior information is available,
the finite sample theory is derived for the proposed test statistic. At
last, the usefulness of the test is illustrated with several applications to
latent variable models widely used in economics and finance. \vskip0.4cm

\end{abstract}

\vskip 2cm \baselineskip=17pt

\section{Introduction}

Latent variable models have been widely used in economics, finance, and many
other disciplines. Two typical models are the dynamic stochastic general
equilibrium models in macroeconomics and stochastic volatility models in
finance. The latent variable models are generally indexed by the latent
variable and the parameter. In many latent variable models, the latent
variable is generally high-dimensional so that the observed likelihood
function which is a marginal integral on the latent variable is often
intractable and becomes difficult to evaluate accurately. Consequently, the
statistical inference for latent variable models is nontrivial in practice.
In the recent years, Bayesian MCMC methods have been applied in more and
more applications in economics and finance due to that they make it possible
to fit increasingly complex models, especially latent variable models, see
Geweke, et al (2011) and reference therein.

In economic research, the point null hypothesis test is a fundamental topic
in statistical inference. Under the Bayesian paradigm, the Bayes factors
(BFs) are the corner-stone of Bayesian hypothesis testing (e.g.
Jeffreys,1961; Kass and Raftery 1995; Geweke, 2007). Unfortunately, the BFs
are not problem-free. First, the BFs are sensitive to the prior distribution
and subjects to the notorious Jeffreys-Lindley's paradox; see for example,
Kass and Raftery (1995), Poirier (1995), Robert (1993, 2001). Second, the
calculation of BFs generally involves the evaluation of marginal likelihood.
In many cases, the evaluation of marginal likelihood is often difficult.

Not surprisingly, some alternative strategies have been proposed to test a
point null hypothesis in the Bayesian literature. In recent years, on the
basis of the statistical decision theory, several interesting Bayesian
approaches to replace BFs have been developed for hypothesis testing. For
example, Bernardo and Rueda (2002, BR hereafter) demonstrated that BFs for
the Bayesian hypothesis testing can be regarded as a decision problem with a
simple zero-one discrete loss function. However, the zero-one discrete
function requires the use of non-regular (not absolutely continuous) prior
and this is why BF leads to Jeffreys-Lindley's paradox. BR further suggested
using a continuous loss function, based on the well-known continuous
Kullback-Leibler (KL) divergence function. As a result, it was shown in BR
that their Bayesian test statistic does not depend on any arbitrary constant
in the prior. However, BR's approach has some disadvantages. First, the
analytical expression of the KL loss function required by BR is not always
available, especially for latent variable models. Second, the test statistic
is not a pivotal quantity. Consequently, BR had to use subjective threshold
values to test the hypothesis.

To deal with the computational problem in BR in latent variable models, Li
and Yu (2012, LY hereafter) developed a new test statistic based on the $%
\mathcal{Q}$ function in the Expectation-Maximization (EM) algorithm. LY
showed that the new statistic is well-defined under improper priors and easy
to compute for latent variable models. Following the idea of McCulloch
(1989), LY proposed to choose the threshold values based on the Bernoulli
distribution. However, like the test statistic proposed by BR, the test
statistic proposed by LY is not pivotal. Moreover, it is not clear if the
test statistic of LY can resolve Jeffreys-Lindley's paradox.

Based on the difference between the deviances, Li, Zeng and Yu (2014, LZY
hereafter) developed another Bayesian test statistic for hypothesis testing.
This test statistic is well-defined under improper priors, free of
Jeffreys-Lindley's paradox, and not difficult to compute. Moreover, its
asymptotic distribution can be derived and one may obtain the threshold
values from the asymptotic distribution. Unfortunately, in general the
asymptotic distribution depends on some unknown population parameters and
hence the test is not pivotal. With sharing the nice properties with Li,
Zeng and Yu (2014, LZY hereafter), Li, Liu and Yu (2015)(2015, LLY
hereafter) further proposed a pivotal Bayesian test statistic, based on a
quadratic loss function, to test a point null hypothesis within the
decision-theoretic framework. However, LLY required to evaluate the first
derivative of the observed log-likelihood. As to the latent variable models,
because the observed log-likelihood is often intractable, this still posed
some tedious computational efforts although there have been several
interesting methods for evaluating the first derivative, such as EM
algorithm, Kalman filter or Particle filter.

In the paper, we want to propose another novel, easy-to-implement Bayesian
statistic for hypothesis testing in the framework of latent variable models.
The new statistic can share the important advantages with LLY. First, it is
well-defined under improper prior distributions and avoids Jeffrey-Lindley's
paradox. Second, under some mild regularity conditions, the statistic is
asymptotically equivalent to the Wald test. Hence, from the large sample
theory, it's asymptotic distribution can be derived to follow the $\chi^2$
distribution so that the threshold values can be easily calibrated from this
distribution. Third, it's statistical error can be derived using the Markov
chain Monte Carlo (MCMC) approach. In addition, most importantly, compared
with the previous test statistics, it is extremely convenient for the latent
variable models. We don't need to evaluate the first-order derivative of the
observed log-likelihood function, which is time consuming and difficult for
the latent variable models. We just need the MCMC output of posterior
simulation. The only effort we should make is the inverse of the posterior
variance matrix of the interest parameter in hypothesis testing.
Fortunately, in most applications, the dimension of the interest parameter
is often not so high that our method can be easily applied. In addition,
when the prior information is available, we establish the finite sample
theoretical properties for the proposed test statistic.

The paper is organized as follows. Section 2 presents the Bayesian analysis
for latent variable models. Section 3 develops the new Bayesian test
statistic from the decisional viewpoint and establishes its finite and large
sample theoretical properties. Section 4 illustrates the new method by using
three real examples in economics and finance. Section 5 concludes the paper.
Appendix collects the proof of all the theoretical results.

\section{Bayesian analysis of latent variable models}

Without loss of generality, let $\mathbf{y}=(\mathbf{y}_{1},\mathbf{y}%
_{2},\cdots ,\mathbf{y}_{n})^{T}$ denote observed variables and ${%
\mbox{\boldmath${z}$}}=({\mbox{\boldmath${z}$}}_{1},{\mbox{\boldmath${z}$}}%
_{2},\cdots ,{\mbox{\boldmath${z}$}}_{n})^{T},$ the latent variables. The
latent variable model is indexed by the parameter, ${\mbox{\boldmath${%
\vartheta}$}}$. Let $p(\mathbf{y}|{\mbox{\boldmath${\vartheta}$}})$ be the
likelihood function of the observed data, and $p(\mathbf{y},{%
\mbox{\boldmath${z}$}}|{\mbox{\boldmath${\vartheta}$}}),$ the complete
likelihood function. The relationship between these two likelihood functions
is:
\begin{equation}
p(\mathbf{y}|{\mbox{\boldmath${\vartheta}$}})=\int p(\mathbf{y},{%
\mbox{\boldmath${z}$}}|{\mbox{\boldmath${\vartheta}$}})d{\mbox{%
\boldmath${z}$}.}  \label{eq01}
\end{equation}%
In many latent variable modes, especially dynamic latent variable models,
the latent variable $\mathbf{z}$ is often dependent on the sample size.
Hence, the integral is high-dimensional and often does not have an
analytical expression so that it is generally very difficult to evaluate.
Consequently, the statistical inferences, such as estimation and hypothesis
testing, are difficult to implement if they are based on the popular maximum
likelihood approach.

In recent years, it has been documented that the latent variables models can
be simply and efficiently analyzed using MCMC techniques under the Bayesian
framework. For details about Bayesian analysis of latent variable models via
MCMC such as algorithms, examples and references, see Geweke, et al. (2011).
Let $p({\mbox{\boldmath${\vartheta}$}})$ be prior distribution of unknown
parameter ${\mbox{\boldmath${\vartheta}$}}$. Owing to the complexity induced
by latent variables, the observed likelihood $p(\mathbf{y}|{%
\mbox{\boldmath${\vartheta}$}})$ is often intractable, hence it is almost
impossible to evaluate the expectation of the posterior density $p({%
\mbox{\boldmath${\vartheta}$}}|\mathbf{y})$ directly. To alleviate this
difficulty, in the posterior analysis, the popular data-augmentation
strategy(Tanner and Wong, 1987) is applied to augment the observed variable $%
\mathbf{y}$ with the latent variable $\mathbf{z}$. Then, the well-known
Gibbs sampler can be used to generate random samples from the joint
posterior distribution $p({\mbox{\boldmath${\vartheta}$}} ,\mathbf{z}|%
\mathbf{y})$. More concretely, we start with an initial value $[{%
\mbox{\boldmath${\vartheta}$}}^{(0)},,\mathbf{z}^{(0)}]$, and then simulates
one by one; at the $j$th iteration, with current values $[{%
\mbox{\boldmath${\vartheta}$}}^{(j)},\mathbf{z}^{(j)}]:$

\begin{description}
\item ~~~~~~~~~~~~~~~(a) Generate ${\mbox{\boldmath${\vartheta}$}}^{(j+1)}$
from $p({\mbox{\boldmath${\vartheta}$}} |\mathbf{z}^{(j)},\mathbf{y})$;

\item ~~~~~~~~~~~~~~~(b) Generate $\mathbf{z}^{(j+1)}$ from $p(\mathbf{z}|{%
\mbox{\boldmath${\vartheta}$}}^{(j+1)},\mathbf{z})$.
\end{description}

After the burning-in phase, that is, sufficiently many iterations of this
iteration procedure, the simulated random samples can be regarded as
efficient random observations from the joint posterior distribution $p({%
\mbox{\boldmath${\vartheta}$}} ,\mathbf{z}|\mathbf{y})$.

The statistical inference can be established on the efficient random
observations drawn from the posterior distribution. Bayesian estimates of ${%
\mbox{\boldmath${\vartheta}$}}$ and latent variables $\mathbf{z}$ as well as
their standard errors can be easily obtained via the corresponding sampling
mean and sample covariance matrix of the generated random observations.
Specifically, let $\{{\mbox{\boldmath${\vartheta}$}}^{(j)},\mathbf{z}%
^{(j)},j=1,2,\cdots, J\}$ be effective random observations generated form
the joint posterior distribution $p({\mbox{\boldmath${\vartheta}$}},\mathbf{z%
}|\mathbf{y})$. Then the joint Bayesian estimates of ${\mbox{\boldmath${%
\vartheta}$}},\mathbf{z}$, as well as the estimates of their covariance
matrix can be obtained as follows:
\begin{eqnarray*}
& &{\mbox{\boldmath${{\widehat\vartheta}}$}}=\frac {1}{J}\sum_{j=1}^J{%
\mbox{\boldmath${\theta}$}}^{(j)},~ \widehat{Var}({\mbox{\boldmath${%
\vartheta}$}}|\mathbf{y})=\frac{1}{J}\sum_{j=1}^J ({\mbox{\boldmath${%
\vartheta}$}}^{(j)}-{\mbox{\boldmath${{\widehat{\vartheta}}}$}})({%
\mbox{\boldmath${\vartheta}$}}^{(j)}-{\mbox{\boldmath${\widehat{\vartheta}}$}%
})^{\prime} \\
& &\mathbf{\widehat{z}}=\frac {1}{J}\sum_{j=1}^J\mathbf{z}^{(j)},~ \widehat{%
Var}(\mathbf{z}|\mathbf{y})=\frac{1}{J}\sum_{j=1}^J (\mathbf{z}^{(j)}-%
\mathbf{\widehat{z}})(\mathbf{z}^{(j)}-\mathbf{\widehat{z}})^{\prime}.
\end{eqnarray*}

\section{Bayesian Hypothesis Testing from the Decision Theory}

\subsection{Testing a point null hypothesis}

It is assumed that a probability model $M\equiv \{p(\mathbf{y}|{%
\mbox{\boldmath${\theta}$}},{\mbox{\boldmath${\psi}$}})\}$ is used to fit
the data. We are concerned with a point null hypothesis testing problem
which may arise from the prediction of a particular theory. Let ${%
\mbox{\boldmath${\theta}$}}\in \mathbf{\Theta }$ denote a vector of $p$%
-dimensional parameters of interest and ${\mbox{\boldmath${\psi}$}}\in
\mathbf{\Psi }$ a vector of $q$-dimensional nuisance parameters. The problem
of testing a point null hypothesis is given by
\begin{equation}
\left\{
\begin{array}{cc}
H_{0}: & {\mbox{\boldmath${\theta}$}}={\mbox{\boldmath${\theta}$}}_{0} \\
H_{1}: & {\mbox{\boldmath${\theta}$}}\neq {\mbox{\boldmath${\theta}$}}_{0}%
\end{array}%
.\right.  \label{eq03}
\end{equation}

The hypothesis testing may be formulated as a decision problem. It is
obvious that the decision space has two statistical decisions, to accept $%
H_{0}$ (name it $d_{0}$) or to reject $H_{0}$ (name it $d_{1}$). Let $\{%
\mathcal{L}[d_{i},({\mbox{\boldmath${\theta}$}},{\mbox{\boldmath${\psi}$}}%
)],i=0,1\}$ be the loss function of statistical decision. Hence, a natural
statistical decision to reject $H_{0}$ can be made when the expected
posterior loss of accepting $H_{0}$ is sufficiently larger than the expected
posterior loss of rejecting $H_{0}$, i.e.,
\begin{equation*}
~~~\mathbf{T}(\mathbf{y},{\mbox{\boldmath${\theta}$}}_{0})=\int_{\Theta
}\int_{\Psi }\left\{ \mathcal{L}[d_{0},({\mbox{\boldmath${\theta}$}},{%
\mbox{\boldmath${\psi}$}})]-\mathcal{L}[d_{1},({\mbox{\boldmath${\theta}$}},{%
\mbox{\boldmath${\psi}$}})]\right\} p({\mbox{\boldmath${\theta}$}},{%
\mbox{\boldmath${\psi}$}}|\mathbf{y})\mbox{d}{\mbox{\boldmath${\theta}$}%
\mbox{d}\mbox{\boldmath${\psi}$}}>c\geq 0,
\end{equation*}%
where $\mathbf{T}(\mathbf{y},{\mbox{\boldmath${\theta}$}}_{0})$ is a
Bayesian test statistic; $p({\mbox{\boldmath${\theta}$}},{%
\mbox{\boldmath${\psi}$}}|\mathbf{y})$ the posterior distribution with some
given prior $p({\mbox{\boldmath${\theta}$}},{\mbox{\boldmath${\psi}$}})$; $c$
a threshold value. Let $\triangle \mathcal{L}[H_{0},({\mbox{\boldmath${%
\theta}$}},{\mbox{\boldmath${\psi}$}})]=\mathcal{L}[d_{0},({%
\mbox{\boldmath${\theta}$}},{\mbox{\boldmath${\psi}$}})]-\mathcal{L}[d_{1},({%
\mbox{\boldmath${\theta}$}},{\mbox{\boldmath${\psi}$}})]$ be the net loss
difference function which can generally be used to measure the evidence
against $H_{0}$ as a function of $({\mbox{\boldmath${\theta}$}},{%
\mbox{\boldmath${\psi}$}})$. Hence, the Bayesian test statistic can be
rewritten as
\begin{equation*}
~~~\mathbf{T}(\mathbf{y},{\mbox{\boldmath${\theta}$}}_{0})=E_{{%
\mbox{\boldmath${\vartheta}$}}|\mathbf{y}}\left( \triangle \mathcal{L}%
[H_{0},({\mbox{\boldmath${\theta}$}},{\mbox{\boldmath${\psi}$}})]\right).
\end{equation*}

\begin{remark}
When the equal prior $p\left( {\mbox{\boldmath${
\theta}$}} ={\mbox{\boldmath${\theta }$}}_{0}\right) =p\left( {%
\mbox{\boldmath${\theta}$}} \neq {\mbox{\boldmath${\theta }$}}_{0}\right) =%
\frac{1}{2}$ and the net loss function is taken as
\begin{equation*}
\Delta \mathcal{L}\left( H_{0},{\mbox{\boldmath${\theta}$}},{%
\mbox{\boldmath${\psi }$}}\right) =%
\begin{cases}
-1 & \text{if }{\mbox{\boldmath${\theta}$}} ={\mbox{\boldmath${\theta }$}}%
_{0} \\
1, & \text{if }{\mbox{\boldmath${\theta}$}} \neq {\mbox{\boldmath${\theta }$}%
}_{0}%
\end{cases}%
\end{equation*}%
following BR (2002) and Li and Yu (2012), the Bayesian test statistic can be
given by
\begin{equation*}
T\left(\mathbf{y},{\mbox{\boldmath${\theta }$}}_{0}\right) =\int_{\Theta
}\int_{\Psi }\Delta \mathcal{L}\left( H_{0},{%
\mbox{\boldmath${\theta ,\psi
}$}}\right) p\left( {\mbox{\boldmath${\theta ,\psi |\mathbf{y}}$}}\right) d{%
\mbox{\boldmath${\theta d\psi}$}}>0
\end{equation*}%
which is equivalent to the well known BFs (Kass and Raftery, 1995) as
\begin{equation*}
BF_{10}=\frac{p(\mathbf{y}|H_1)}{p(\mathbf{y}|H_0)}=\frac{\int p(\mathbf{y},%
\mathbf{h},{\mbox{\boldmath${\vartheta}$}})\mbox{d}\mathbf{h}\mbox{d}{%
\mbox{\boldmath${\vartheta}$}}}{\int p(\mathbf{y},\mathbf{h},{%
\mbox{\boldmath${{\psi}}$}}|{\mbox{\boldmath${\theta}$}}_0)\mbox{d}\mathbf{h}%
\mbox{d}{\mbox{\boldmath${{\psi}}$}}}>1
\end{equation*}
when rejecting the null hypothesis. In practice, the BFs are often served as
the gold statistics for hypothesis testing and the benchmark for the other
test statistics. However, the BFs have some theoretical and computational
difficulties. First, in the literature, it is well documented that it can
not be well defined when using improper priors and suffers from the
notorious Jeffreys-Lindley's paradox, see Poirier (1995), Robert (2001), Li
and Yu (2012), Li, Zeng and Yu (2014), etc. Second, the computation of $%
BF_{10}$ requires to evaluate the marginal likelihood $p(\mathbf{y}|H_i),
i=0,1$. Clearly, for latent variable models, this often involves a
marginalization over the unknown latent variables $\mathbf{h}$ and the
parameter ${\mbox{\boldmath${\vartheta}$}}$. Furthermore, it is often a
high-dimensional integration and generally hard to do in practice although
there have been several interesting methods proposed in the literature for
computing BFs from the MCMC output; see, for example, Chib (1995), and Chib
and Jeliazkov (2001).
\end{remark}

\begin{remark}
Under decision theory framework, several papers have explored some effective
approaches to replace the BFs for point-null hypothesis testing. Poirier
(1997) developed a loss function approach for hypothesis testing for models
without latent variables. Bernardo and Rueda (2002) proposed an intrinsic
statistic for Bayesian hypothesis test based on the Kullback-Leibler (KL)
loss function. However, the analytical expression of the KL loss function
required by BR is not always available, especially for latent variable
models. Furthermore, the test statistic is not a pivotal quantity so that BR
had to use subjective threshold values for hypothesis testing. To deal with
latent variable models, Li and Yu (2012) proposed a Bayesian test statistic
based on the Q-function loss function within EM algorithm. LY showed that
the test statistic is well-defined under improper priors and easy to compute
for latent variable models. However, like the test statistic proposed by BR,
the test statistic proposed by LY is not pivotal. Moreover, it is not clear
if the test statistic of LY can resolve Jeffreys-Lindley's paradox. Li, Zeng
and Yu (2014) proposed another test statistic, which is a Bayesian version
of likelihood ratio test statistic. This test statistic is well-defined
under improper priors, free of Jeffreys-Lindley's paradox, and not difficult
to compute. Moreover, its asymptotic distribution can be derived and one may
obtain the threshold values from the asymptotic distribution. Unfortunately,
in general the asymptotic distribution depends on some unknown population
parameters and hence the test is not pivotal.
\end{remark}

\begin{remark}
In a recent paper, Li, Liu and Yu (2015) proposed a new Bayesian test
statistic with the following quadratic loss function
\begin{equation*}
\Delta l\left( H_{0},{\mbox{\boldmath${\theta ,\psi }$}}\right) =\left( {%
\mbox{\boldmath${\theta }$}}-\bar{{\mbox{\boldmath${\theta
}$}}}\right) ^{\prime }C_{\theta \theta }\left( \bar{{\mbox{\boldmath${%
\vartheta }$}}}_{0}\right) \left( {\mbox{\boldmath${\theta }$}}-\bar{{%
\mbox{\boldmath${\theta }$}}}\right) ,
\end{equation*}%
where $\bar{{\mbox{\boldmath${\vartheta }$}}}_{0}=\left( {%
\mbox{\boldmath${\theta }$}}_{0},\bar{{\mbox{\boldmath${\psi }$}}}%
_{0}\right) $ is the posterior mean under the null and $C_{\theta \theta
}\left( {\mbox{\boldmath${\vartheta }$}}\right) $ is the submatrix of $%
C\left( {\mbox{\boldmath${\vartheta }$}}\right) =\left\{ \frac{\partial \log
p\left( \mathbf{y},{\mbox{\boldmath${\vartheta}$}} \right) }{\partial {%
\mbox{\boldmath${\vartheta }$}}}\right\} \left\{ \frac{\partial \log p\left(%
\mathbf{y},{\mbox{\boldmath${\vartheta}$}}\right) }{\partial {%
\mbox{\boldmath${\vartheta }$}}}\right\} ^{\prime }$ with respect to
parameters ${\mbox{\boldmath${\theta }$}}$. With this loss function, they
showed that under some mild regularity conditions, the proposed Bayesian
test statistics followed a pivotal $\chi _{p}^{2}$ asymptotically, hence, it
is very easy to calibrate threshold values. Furthermore, this proposed test
statistic shared some nice properties with Li and Yu (2012), Li,Zeng and Yu
(2014), that is, this test statistic is well-defined under improper prior
and immune to Jefferys-Lindley's paradox. As to latent variable models,
obviously, the test statistic by Li, Liu and Yu (2015) needs to evaluate the
first-derivative of the observed likelihood function. As noted in section 2,
the observed likelihood function often generally doesn't have analytical
form so that it is not easy to do. Li, Liu and Yu (2015) showed that some
complex simulation algorithms such as EM algorithm, Kalman filter, Particle
filter have to be applied for evaluating the first derivative. \textbf{%
Further, the standard error of the new statistic will be smaller than the
one in LLY.}
\end{remark}

\subsection{A new Bayesian $\protect\chi^2$ test from decision theory}

In this subsection, as to latent variable models, based on the decision
theory, we develop a new Bayesian $\chi^2$ test statistic for hypothesis
testing. The new test statistic can share the nice advantages with Li, Liu
and Yu (2015). For example, it can be well-defined under improper prior
distributions and avoids Jeffrey-Lindley's paradox. Furthermore, the
threshold values can be easily calibrated from the pivotal asymptotic
distribution and it's statistical error can be derived using MCMC approach.
Most importantly, the new test statistic can achieve other important
advantages over the existing approaches, such as, Li, et al (2015). Our new
contributions are twofold. As to latent variable models, it can be shown
that the new test statistic is only the by-product of the posterior outputs,
hence, very easy to compute. In addition, when the prior information is
available, we establish the finite sample theory.

As to any ${\mbox{\boldmath${\tilde\vartheta}$}}$ in support space of ${%
\mbox{\boldmath${\vartheta}$}}$, let
\begin{equation*}
\mathbf{V}({\mbox{\boldmath${\tilde\vartheta}$}})=E\left[({%
\mbox{\boldmath${\vartheta}$}}-{\mbox{\boldmath${\tilde\vartheta}$}})({%
\mbox{\boldmath${\vartheta}$}}-{\mbox{\boldmath${\tilde\vartheta}$}}%
)^{\prime}|\mathbf{y},H_1\right]=\int ({\mbox{\boldmath${\vartheta}$}}-{%
\mbox{\boldmath${\tilde\vartheta}$}})({\mbox{\boldmath${\vartheta}$}}-{%
\mbox{\boldmath${\tilde\vartheta}$}})^{\prime}p({\mbox{\boldmath${%
\vartheta}$}}|\mathbf{y})\mbox{d}{\mbox{\boldmath${\vartheta}$}}
\end{equation*}
In this paper, under the statistical decision theory, we propose the
following net loss function for hypothesis testing
\begin{equation*}
\triangle \mathcal{L}[H_{0},({\mbox{\boldmath${\theta}$}},{%
\mbox{\boldmath${\psi}$}})]=\left({\mbox{\boldmath${\theta}$}}-{%
\mbox{\boldmath${\theta}$}}_{0}\right)^{\prime}\left[\mathbf{V}%
_{\theta\theta}({\mbox{\boldmath${\bar\vartheta}$}})\right]^{-1}\left({%
\mbox{\boldmath${\theta}$}}-{\mbox{\boldmath${\theta}$}}_{0}\right)
\end{equation*}%
where $\mathbf{V}_{\theta\theta}({\mbox{\boldmath${\bar\vartheta}$}})$ is
the submatrix of $\mathbf{V}({\mbox{\boldmath${\bar\vartheta}$}})$
corresponding to ${\mbox{\boldmath${\theta}$}}$, $\left[\mathbf{V}%
_{\theta\theta}({\mbox{\boldmath${\bar\vartheta}$}})\right]^{-1}$ is the
inverse matrix of $\mathbf{V}_{\theta\theta}({\mbox{\boldmath${\bar%
\vartheta}$}})$ and ${\mbox{\boldmath${\bar\vartheta}$}}$ is the posterior
mean of ${\mbox{\boldmath${\vartheta}$}}$ under the alternative hypothesis $%
H_1$. Then, we can define a Bayesian test statistic as follows:
\begin{equation}
\mathbf{T}(\mathbf{y},{\mbox{\boldmath${\theta}$}}_0)=\int \triangle
\mathcal{L}[H_{0},({\mbox{\boldmath${\theta}$}},{\mbox{\boldmath${\psi}$}}%
)]p({\mbox{\boldmath${\vartheta}$}}|\mathbf{y})\mbox{d}{\mbox{\boldmath${%
\vartheta}$}}=\int \left({\mbox{\boldmath${\theta}$}}-{\mbox{\boldmath${%
\theta}$}}_{0}\right)^{\prime}\left[\mathbf{V}_{\theta\theta}({%
\mbox{\boldmath${\bar\vartheta}$}})\right]^{-1}({\mbox{\boldmath${\bar%
\vartheta}$}}) \left({\mbox{\boldmath${\theta}$}}-{\mbox{\boldmath${\theta}$}%
}_{0}\right)p({\mbox{\boldmath${\vartheta}$}}|\mathbf{y})\mbox{d}{%
\mbox{\boldmath${\vartheta}$}}
\end{equation}

\begin{remark}
When informative priors are not available, an objective prior or default
prior may be used. Often, $p ({\mbox{\boldmath${\theta}$}})$ is taken as
uninformative priors, such as Jeffreys or the reference prior (Jeffreys,
1961; Berger and Bernardo, 1992). These priors are generally improper, and
it follows that $p({\mbox{\boldmath${\vartheta}$}})=Af({\mbox{\boldmath${%
\vartheta}$}})$ where $f({\mbox{\boldmath${\vartheta}$}})$ is a
nonintegrable function, and $A$ is an arbitrary positive constant. Since the
posterior distribution $p({\mbox{\boldmath${\vartheta}$}}|\mathbf{y})$ is
independent of an arbitrary constant in the prior distributions, and $%
\mathbf{V}_{\theta\theta}({\mbox{\boldmath${\bar\vartheta}$}})$ is the
posterior covariance matrix of the interest parameter ${\mbox{\boldmath${%
\theta}$}}$, hence, the statistic is independent of an arbitrary constant.
Consequently, our proposed test statistic $\mathbf{T}({\mbox{\boldmath${%
\mathbf{y}}$}},{\mbox{\boldmath${\theta_0}$}})$ is independent on this
arbitrary positive constant and can be well-defined under improper priors.
\end{remark}

\begin{remark}
To see how the new statistic can avoid Jeffreys-Lindley's paradox, consider
the example discussed in Li, et al (2015). Let $y_1,y_2,\cdots,y_n\sim
N(\theta ,\sigma ^{2})$ with a known $\sigma ^{2}$ and we test the null
hypothesis $H_{0}:\theta =0$. Let the prior distribution of $\theta $ be $%
N(\mu ,\tau ^{2})$. The prior distribution of $\theta $ can be set as $%
N(\mu_0,\tau ^{2})$ with $\mu_0=0$. Suppose $\mathbf{y}=(y_{1},...,y_{n}),
\bar{y}=\frac{1}{n}\sum_{i=1}^{n}y_{i}$. We want to test the simple point
null hypothesis $H_{0}:\theta =0$. The posterior distribution of $\theta $
is $N(\mu(\mathbf{y}),\omega ^{2})$ with
\begin{eqnarray*}
\mu(\mathbf{y})=\frac{n\tau ^{2}\bar{y}}{\sigma ^{2}+n\tau ^{2}},\omega ^{2}=%
\frac{\sigma ^{2}\tau ^{2}}{\sigma ^{2}+n\tau ^{2}},
\end{eqnarray*}

It can be shown that
\begin{eqnarray*}
& &2\log BF_{10}=\frac{n\tau^2}{n\tau^2+\sigma^2}\frac{n\bar{y}^2}{\sigma^2}%
+\log\frac{\sigma^2} {n\tau ^{2}+\sigma^2} \\
& &\mathbf{T}(\mathbf{y}, \theta_{0})=\frac{n\tau^2}{n\tau^2+\sigma^2}\frac{n%
\bar{y}^2}{\sigma^2}+1
\end{eqnarray*}
Clearly, when the prior information is very uninformative, as $%
\tau^2\rightarrow +\infty$, we can get that $\log BF_{10}\rightarrow-\infty$
which means that the BFs always support the null hypothesis. This is
well-known as Jeffreys-Lindley's paradox in the Bayesian literature.
However, we can find that $\mathbf{T}(\mathbf{y},\theta_0)\rightarrow \frac{n%
\bar{y}^2}{\sigma^2}+1$ as $\tau^2\rightarrow +\infty$. Hence, $\mathbf{T}(%
\mathbf{y},\theta_0)$ is distributed as $\chi ^{2}(1)+1$ when $H_{0}$ is
true. Consequently, our proposed test statistic is immune to
Jeffreys-Lindley's paradox.
\end{remark}

\begin{remark}
The implementation of the Bayesian test statistic by Li,et al (2015)
requires the evaluation of the first derivative of the observed
log-likelihood function. As described in section 2, for latent variable
models, the observed likelihood function generally doesn't have analytical
form so that it is generally hard to get the fist derivative. Compared with
Li,et al (2015), the main advantage of the proposed test statistic in this
paper is that it is not highly computational intensive. From the equation
(3), we can easily observe that it doesn't require to evaluate the first
derivatives. From the computational perspective, our test statistic is only
involved of the posterior random samples and the inverse of the posterior
covariance matrix. In practice, through the latent variable $\mathbf{z}$ or
parameter ${\mbox{\boldmath${\vartheta}$}}$ may be high-dimensional, in
manly latent variable models, the interest parameter ${\mbox{\boldmath${%
\theta}$}}$ is often low-dimensional. Hence, the proposed Bayesian test
statistic is only by-product of Bayesian posterior output, not requires
additional computational efforts. This is especially advantageous for latent
variable models.
\end{remark}

\subsection{Large sample theory for the Bayesian test statistic}

In this subsection, we establish the Bayesian large sample theory for the
proposed test statistic. Let $\left\{ z_{t}\right\} $ be a sequence of
random vectors defined on the probability space $(\Omega ,\mathcal{F},P)$
and $z^{t}$ be the collection of $\left( z_{1},z_{2},\ldots ,z_{t}\right) $.
Let $y_{t}$ denote an element of $z_{t}$ and write $z_{t}$ as $%
(y_{t},w_{t}^{\prime })^{\prime }$, then we can write the conditional
likelihood function for $y_{t}$ as $f_{t}\left( y_{t}|x_{t},\vartheta
\right) $, where $x_{t}$ include some elements of $w_{t}$ and $z^{t-1}$.
Define $g_{t}\left( \vartheta \right) =g_{t}\left( z^{t},\vartheta \right)
=\log f_{t}\left( y_{t}|x_{t},\vartheta \right) $ to be the conditional
likelihood for $t$ observation and $\nabla ^{j}g_{t}\left( \vartheta \right)
$ as the $jth$ derivative of $g_{t}\left( \vartheta \right) $, we suppress
the subscript when $j=1$. The logarithm of posterior likelihood function is
\begin{equation*}
\mathcal{L}_{n}({\mbox{\boldmath${\vartheta}$}})=\log p({\mbox{\boldmath${%
\vartheta}$}}|\mathbf{y}).
\end{equation*}%
Furthermore, let $\dot{\mathcal{L}}_{n}({\mbox{\boldmath${\vartheta}$}}%
)=\partial \log p({\mbox{\boldmath${\vartheta}$}}|\mathbf{y})/\partial {%
\mbox{\boldmath${\vartheta}$}}$, $\ddot{\mathcal{L}}_{n}({%
\mbox{\boldmath${\vartheta}$}})=\partial ^{2}\log p({\mbox{\boldmath${%
\vartheta}$}}|\mathbf{y})/\partial {\mbox{\boldmath${\vartheta}$}}{\partial %
\mbox{\boldmath${\vartheta}$}}^{\prime }$ and the negative Hessian matrix as
\begin{equation*}
\mathbf{I}({\mbox{\boldmath${\vartheta}$}})=-\frac{\partial ^{2}\log p(%
\mathbf{y}|{\mbox{\boldmath${\vartheta}$}})}{\partial {\mbox{\boldmath${%
\vartheta}$}}\partial {\mbox{\boldmath${\vartheta}$}}^{\prime }}.
\end{equation*}%
Let the prior density to be $p({\mbox{\boldmath${\vartheta}$}})$, $\gamma({%
\mbox{\boldmath${\vartheta}$}})=\log p({\mbox{\boldmath${\vartheta}$}})$ and
$\gamma^{{\mbox{\boldmath${\vartheta}$}}}({\mbox{\boldmath${\vartheta}$}}%
)=\partial\log p({\mbox{\boldmath${\vartheta}$}})/\partial{%
\mbox{\boldmath${\vartheta}$}}$. In order to derive the asymptotic
distribution of the proposed test statistic, following LZY (2014) and
LLY(2015), a set of regularity conditions are imposed in the following.


\begin{assumption}
There exists a finite sample size $n^{\ast }$, so that, for $n>n^{\ast }$,
there is a local maximum at ${\mbox{\boldmath${\widehat\vartheta_m}$}}$
(i.e., posterior mode) such that $\dot{\mathcal{L}}_{n}({\mbox{\boldmath${%
\widehat\vartheta_m}$}})=0$ and $\ddot{\mathcal{L}}_{n}({\mbox{\boldmath${%
\widehat\vartheta_m}$}})$ is negative definite.
\end{assumption}


\begin{assumption}
The largest eigenvalue $\lambda _{n}$ of $-\ddot{\mathcal{L}}_{n}^{-2}(%
\widehat{\mbox{\boldmath${\vartheta}$}}_{m})$ goes to zero in probability as
$n\rightarrow \infty $.
\end{assumption}


\begin{assumption}
For any $\varepsilon >0$, there exists a positive number $\delta $, such that%
\begin{equation}
\lim_{n\rightarrow \infty }P\left[ \sup_{{\mbox{\boldmath${\vartheta}$}} \in
B\left( {\mbox{\boldmath${\widehat\vartheta_m}$}},\text{ }\delta \right)
}\left\Vert \ddot{\mathcal{L}}^{-1}\left( {\mbox{\boldmath${\widehat%
\vartheta_m}$}}\right) \left[ \ddot{\mathcal{L}}\left( {\mbox{\boldmath${%
\widehat\vartheta}$}} \right) -\ddot{\mathcal{L}}\left( {\mbox{\boldmath${%
\widehat\vartheta_m}$}}\right) \right] \right\Vert <\varepsilon \right] =1 .
\label{A3}
\end{equation}
\end{assumption}


\begin{assumption}
For any $\delta>0$,
\begin{equation*}
\int_{{\mbox{\boldmath${\Omega}$}}-B(\widehat{\mbox{\boldmath${\vartheta}$}}%
_{m},\delta )}p({\mbox{\boldmath${\vartheta}$}}|\mathbf{y})d{{%
\mbox{\boldmath${\vartheta}$}}}\rightarrow 0,
\end{equation*}%
in probability as $n\rightarrow \infty $, where $\mathbf{\Omega }$ is the
support space of ${\mbox{\boldmath${\vartheta}$}}$.
\end{assumption}


\begin{assumption}
For any $\delta>0$,
\begin{equation*}
\int_{{\mbox{\boldmath${\Omega}$}}-B(\widehat{\mbox{\boldmath${\vartheta}$}}%
_{m},\delta )}\left\Vert {\mbox{\boldmath${\vartheta}$}} \right\Vert ^{2}p({%
\mbox{\boldmath${\vartheta}$}}|\mathbf{y})d{{\mbox{\boldmath${\vartheta}$}}}%
= O_p(n^{-3}),
\end{equation*}%
as $n\rightarrow \infty $, where $\mathbf{\Omega }$ is the support space of $%
{\mbox{\boldmath${\vartheta}$}}$.
\end{assumption}


\begin{assumption}
Let $\mathcal{\vartheta }_{0}$ to be true value, $\mathcal{\vartheta }%
_{0}\in int\left( \Theta \right) $ where $\Theta $ is a compact, separable
metric space.
\end{assumption}


\begin{assumption}
$\left\{w_{t},t=1,2,3,\ldots \right\} $ is an $\alpha$ mixing sequence that
satisfies, for $\mathcal{F}_{-\infty }^{t}=\sigma \left(
z_{t},z_{t-1},\ldots \right) $ and $\mathcal{F}_{t+m}^{\infty }=\sigma
\left( z_{t+m},z_{t+m+1},\ldots \right) $, the mixing coefficient $\alpha
\left( m\right) =O\left( m^{\frac{-r}{r-2}-\varepsilon }\right) $ for some $%
\varepsilon >0$ and $r>2$.
\end{assumption}


\begin{assumption}
Let $N_{\delta }\left( \mathcal{\vartheta }_{\ast }\right) =\left\{ \mathcal{%
\vartheta \in }\Theta :\left\Vert \mathcal{\vartheta -\vartheta }_{\ast
}\right\Vert \leq \delta \right\} $ for $\mathcal{\vartheta }_{\ast }\in
\Theta $, $\delta \geq 0$ and $0\leq j \leq s_{1}$, (i) $\sup_{\mathcal{%
\vartheta \in }N_{\delta }\left( \mathcal{\vartheta }_{\ast }\right)
}\nabla^{j}g_{t}\left( \mathcal{\vartheta }\right) $ and $\inf_{\mathcal{%
\vartheta \in }N_{\delta }\left( \mathcal{\vartheta }_{\ast }\right)
}\nabla^{j}g_{t}\left( \mathcal{\vartheta }\right) $ are measurable to $%
\mathcal{F}_{-\infty }^{t}$ and strictly stationary; (ii) $E\left[ \sup_{%
\mathcal{\vartheta \in }N_{\delta }\left( \mathcal{\vartheta }_{\ast
}\right) }\nabla^{j}g_{t}\left( \mathcal{\vartheta }\right) \right] <\infty $
and $E\left[ \inf_{\mathcal{\vartheta \in }N_{\delta }\left( \mathcal{%
\vartheta }_{\ast }\right) }\nabla^{j}g_{t}\left( \mathcal{\vartheta }%
\right) \right] >-\infty $; (iii) $\lim_{\delta \downarrow 0}E\left[ \sup_{%
\mathcal{\vartheta \in }N_{\delta }\left( \mathcal{\vartheta }_{\ast
}\right) }\nabla^{j}g_{t}\left( \mathcal{\vartheta }\right) \right]
=\lim_{\delta \downarrow 0}E\left[ \inf_{\mathcal{\vartheta \in }N_{\delta
}\left( \mathcal{\vartheta }_{\ast }\right) }\nabla^{j}g_{t}\left( \mathcal{%
\vartheta }\right) \right] =E\left[ \nabla^{j}g_{t}\left( \mathcal{\vartheta
}_{\ast }\right) \right] $.
\end{assumption}


\begin{assumption}
There exists a function $M_{t}(\omega _{t})$ such that for $0\leqslant j
\leqslant s_{2} $, all $\theta \in \mathcal{G}$ where $\mathcal{G}$ is an
open, convex set containing $\Theta $, $\bigtriangledown ^{j}g_{t}\left(
\vartheta \right) $ exists, $\sup_{t}\left\Vert M_{t}(\omega
_{t})\right\Vert ^{r+\delta }\leq M<\infty $ for some $\delta >0$.
\end{assumption}


\begin{assumption}
The prior density is continuous and $0<p(\vartheta)<\infty$ for all $%
\vartheta\in\Theta$.
\end{assumption}


\begin{assumption}
For $0<j<s_{3}$, $E\left\Vert\bigtriangledown ^{j}\gamma\left( \vartheta
\right)\right\Vert=O(1) $ .
\end{assumption}

\begin{remark}
Regularity Assumptions 1-4 have been used to develop the Bayesian large
sample theory. This theory is proved by chen(1985), which states that the
posterior distribution is degenerate about the posterior mode and
asymptotically normal after suitable scaling, that is,
\begin{eqnarray*}
\left({\mbox{\boldmath${\vartheta}$}}-\widehat{\mbox{\boldmath${\vartheta}$}}%
_{m}\right)|\mathbf{y}\overset{d}{\longrightarrow} N\left[0, -\ddot{\mathcal{%
L}}_{n}^{-1}(\widehat{\mbox{\boldmath${\vartheta}$}}_{m})\right]
\end{eqnarray*}
More details, one can refer to Chen (1985). Bickel and Doksum (2006), and Le
Cam and Yang (2000), Ghosh (2003) presented another version of this theorem
on the basis of other similar regularity conditions. The main difference is
that the value at which the asymptotic posterior variance matrix is
evaluated. It is the posterior mode $\widehat{\mbox{\boldmath${\vartheta}$}}%
_{m}$ in Chen (1985), the true value $\widehat{\mbox{\boldmath${\vartheta}$}}%
_{0}$ in Bickel and Doksum (2006), and Le Cam and Yang (2000), Ghosh (2003)
and the MLE estimator $\widehat{{\mbox{\boldmath${\vartheta}$}}}$ in Kim
(1994) depending on different assumptions.
\end{remark}

\begin{remark}
Under Assumptions 1-4, conditional on the observed data $\mathbf{y}$, it can
be shown that
\begin{eqnarray*}
&&\bar{{\mbox{\boldmath${\vartheta}$}}}=E\left[ {\mbox{\boldmath${%
\vartheta}$}}|\mathbf{y},H_1\right] =\widehat{\mbox{\boldmath${\vartheta}$}}%
_{m}+o_p(n^{-1/2}), \\
&&\mathbf{V}\left(\widehat{\mbox{\boldmath${\vartheta}$}}_{m}\right) =E\left[
\left({\mbox{\boldmath${\vartheta}$}}-\widehat{\mbox{\boldmath${\vartheta}$}}%
_{m}\right) \left({\mbox{\boldmath${\vartheta}$}}-\widehat{%
\mbox{\boldmath${\vartheta}$}}_{m}\right) ^{^{\prime }}|\mathbf{y},H_1\right]
=-\ddot{\mathcal{L}}^{-1}_{n}\left(\widehat{{\mbox{\boldmath${\vartheta}$}}}%
_{m}\right) +o_p(n^{-1}),
\end{eqnarray*}
where $\bar{{\mbox{\boldmath${\vartheta}$}}}$ is the posterior mean. These
conclusions have been given by Li,Zeng and Yu (2014).
\end{remark}

\begin{remark}
Following Rilstone, Srivatsava and Ullah(1996), Bester and Hansen (2006),
the assumptions 5-10 are used to justify the validity of high order Laplace
expansion. The assumption 5 is analogous to the analytical assumptions for
Laplace's method (kass et al., 1990), but we impose the higher order
constraints $O_p(n^{-3})$ other than $O_p(n^{-2})$, see also Miyata (2004,
2010). With these assumptions, we can get the standard form higher order
Laplace expansion of the order $O_p(n^{-2})$ in kass et al. (1990) to $%
O_p(n^{-3})$, similar to the fully exponential form in Miyata (2004, 2010).
\end{remark}

Let $\widehat{{\mbox{\boldmath${\vartheta}$}}}$ to be the maximum likelihood
estimator of ${\mbox{\boldmath${\vartheta}$}}$ and $\widehat{{%
\mbox{\boldmath${\theta}$}}}$ is the subvector of $\widehat{{%
\mbox{\boldmath${\vartheta}$}}}$ corresponding to ${{\mbox{\boldmath${%
\theta}$}}}$, under Assumptions 5-8 with $s_{1}=2$ and $s_{2}=2$, the Wald
statistic be
\begin{equation*}
\mathbf{Wald}=\left(\widehat{{\mbox{\boldmath${\theta}$}}}-{{%
\mbox{\boldmath${\theta}$}}}_0\right)^{\prime}\left[-\ddot{\mathcal{L}}_{n,\theta\theta}^{-1}(\widehat{\mbf{\vartheta}})\right]%
^{-1} \left(\widehat{{\mbox{\boldmath${\theta}$}}}-{{\mbox{\boldmath${%
\theta}$}}}_0\right),
\end{equation*}%
where $\ddot{\mathcal{L}}_{n,\theta\theta}^{-1}(\widehat{\mbf{\vartheta}})$ is the submatrix of $\ddot{\mathcal{L}}_{n}^{-1}(\widehat{\mbf{\vartheta}})$
corresponding to ${\mbox{\boldmath${\theta}$}}$.

\begin{theorem}
\label{thm1} Under Assumptions 6-11 with $s_{1}=s_{2}=s_{3}=3$, when the
likelihood dominates the prior such as $p({\mbox{\boldmath${\vartheta}$}}%
)=O_p(1)$, under the null hypothesis, we can show that
\begin{eqnarray}
&&\mathbf{T}({\mbox{\boldmath${\mathbf{y}}$}},{\mbox{\boldmath${\theta}$}}%
_{0})-p=\mathbf{Wald}+o_{p}(1)\overset{d}{\rightarrow}\chi^2(p)
\end{eqnarray}
\end{theorem}

\begin{remark}
In Theorem \ref{thm1}, we can see that under the null hypothesis, the
asymptotic distribution of $\mathbf{T}(\mathbf{y},{\mbox{\boldmath${\theta}$}%
}_{0})$ always follows the $\chi ^{2}$ distribution, hence, is pivotal. As
to the proposed test statistic, we still need to specify some threshold
values, $c$ for implementing the test, that is,
\begin{equation*}
\text{Accept}~~H_{0}\text{ if }\mathbf{T}(\mathbf{y},{\mbox{\boldmath${%
\theta}$}}_{0})\leq c;\text{ Reject}~~H_{0}\ \text{ if }\mathbf{T}(\mathbf{y}%
,{\mbox{\boldmath${\theta}$}}_{0})>c.
\end{equation*}
Hence, this asymptotic $\chi^2$ distribution can be utilized conveniently to
calibrate threshold values.
\end{remark}

\begin{remark}
From this theorem, $\mathbf{T}(\mathbf{y},{\mbox{\boldmath${\theta}$}}_{0})$
may be regarded as the Bayesian version of the $\mathbf{Wald}$ statistic.
However, the $\mathbf{Wald}$ test statistic is a frequentist test which is
based on the maximum likelihood estimation of the model in the alternative
hypothesis whereas our test is a Bayesian test which is based on the
posterior quantities of the models under the alternative hypothesis.
\end{remark}

\begin{remark}
The implementation of the $\mathbf{Wald}$ test requires the ML estimation of
the model and evaluation of the second derivative of the observed likelihood
function under the alternative hypothesis. As described in section 2, for
latent variable models, the observed likelihood function generally doesn't
have analytical form so that it is generally hard to get the maximum
likelihood estimator and its corresponding second derivative. Hence, it is
difficult to apply the $\mathbf{Wald}$ test statistic for hypothesis
testing. However, our proposed Bayesian test statistic is only by-product of
posterior outputs. As long as the Bayesian MCMC methods are applicable, our
test can be implemented for latent variable models. In addition, from
equation (3), $\mathbf{T}({\mbox{\boldmath${\mathbf{y}}$}},{%
\mbox{\boldmath${\theta_0}$}})$ can incorporate the prior information
through the posterior distribution directly, but $\mathbf{Wald}$ can not
incorporate the useful prior information.
\end{remark}

\begin{remark}
We use a simple example to illustrate the influence of the prior
distributions. Let $y_{1},...,y_{n}\sim N(\theta ,\sigma ^{2})$ with a known
variance $\sigma ^{2}=1$. The true value of $\theta $ is set at $\theta
_{0}=0.10$. The prior distribution of $\theta $ is set as $N(\mu _{0},\tau
^{2})$. The simple point null hypothesis is $H_{0}:\theta =0$. It can be
shown that
\begin{eqnarray*}
&&2\log BF_{10}=\frac{\sigma ^{2}\tau^2}{\sigma ^{2}+n\tau ^{2}}\left(\frac{n%
\bar{y}}{\sigma^{2}}+\frac{\mu_0}{\tau^2}\right)^2+\log \frac{\sigma ^{2}}{%
\sigma ^{2}+n\tau ^{2}}, \\
&&\mathbf{T}(\mathbf{y},\theta _{0})=\frac{\sigma ^{2}\tau^2}{\sigma
^{2}+n\tau ^{2}}\left(\frac{n\bar{y}}{\sigma^{2}}+\frac{\mu_0}{\tau^2}%
\right)^2+1,\mathbf{Wald}=\frac{n\bar{y}^{2}}{\sigma ^{2}},
\end{eqnarray*}%
where $\bar{y}=\frac{1}{n}\sum_{i=1}^{n}y_{i}$. When $n\longrightarrow
\infty $, $\mathbf{T}(\mathbf{y},\theta _{0})-1\longrightarrow $ $\mathbf{%
Wald}$ and the asymptotic distribution for both $\mathbf{T}(\mathbf{y}%
,\theta _{0})-1$ and $\mathbf{Wald}$ is $\chi ^{2}(1)$. Let us consider the
case that corresponds to an informative prior $N(0.10,10^{-3})$ and compare
it to the case that corresponds to a non-informative prior $N(0,10^{50})$.
Table 1 reports $2\log BF_{10}$, $\mathbf{T}(\mathbf{y},\theta _{0})$, and $%
\mathbf{Wald}$ when $n=10,100,1000,10000$ under these two priors. It can be
seen that both the BF and the new test depend on the prior (although the BFs
tend to choose the wrong model under the vague prior even when the sample
size is very large) while the $\mathbf{Wald}$ test is independent of the
prior. When $n=10,100$, $\mathbf{T}(\mathbf{y},\theta _{0}) $ correctly
rejects the null hypothesis when the prior is informative but fails to
reject it when the prior is vague under the significant level 5\%. In this
case, the $\mathbf{Wald}$ test fails to reject the null hypothesis under
both priors.\footnote{%
To implement the $\mathbf{Wald}$ test,we use the following Fisher's scale.
Let $\alpha $ be the critical level and $P=1-\alpha $. If $P$ is between
95\% and 97.5\%, the evidence for the alternative is \textquotedblleft
moderate\textquotedblright ; between 97.5\% and 99\%, \textquotedblleft
substantial\textquotedblright ; between 99\% and 99.5\%, \textquotedblleft
strong\textquotedblright ; between 99.5\% and 99.9\%, \textquotedblleft very
strong\textquotedblright ; larger than 99.9\%, \textquotedblleft
overwhelming\textquotedblright . To implement the BF we use Jeffreys' scale
instead. If $\log BF_{10}$ is less than 0, there is \textquotedblleft
negative\textquotedblright\ evidence for the alternative; between 0 and 1,
\textquotedblleft not worth more than a bare mention\textquotedblright ;
between 1 and 3, \textquotedblleft positive\textquotedblright ; between 3
and 5, \textquotedblleft strong\textquotedblright ; larger than 5,
\textquotedblleft very strong\textquotedblright .}
\begin{table}[tbph]
\caption{Comparison of $2\log BF_{10}$, $\mathbf{T}(\mathbf{y},\protect%
\theta _{0})$, and $\mathbf{Wald}$}
\begin{center}
\vspace{1em}
\begin{tabular}{c|cccc|cccc}
\hline
Prior & \multicolumn{4}{|c|}{$N(0.10,10^{-3})$} & \multicolumn{4}{c}{$%
N(0,10^{50})$} \\ \hline
$n$ & 10 & 100 & 1000 & 10000 & 10 & 100 & 1000 & 10000 \\ \hline
$2\log BF_{10}$ & 9.96 & 11.12 & 20.60 & 93.58 & -117.42 & -118.50 & -110.72
& -38.00 \\ \hline
$\mathbf{T}(\mathbf{y},\theta _{0})$ & 10.96 & 12.22 & 22.30 & 96.98 & 1.01
& 2.23 & 12.32 & 87.03 \\ \hline
$\mathbf{Wald}$ & 0.01 & 1.23 & 11.32 & 86.03 & 0.01 & 1.23 & 11.32 & 86.03
\\ \hline
\end{tabular}%
\end{center}
\end{table}
\end{remark}

\begin{remark}
Under the null hypothesis, our statistic can be written as
\begin{equation*}
\mathbf{T}(\mathbf{y},\theta _{0})=1+n\frac{\overline{y}^{2}}{\sigma ^{2}}+2%
\overline{y}\frac{\mu _{0}}{\tau ^{2}}-\overline{y}^{2}\frac{1}{\tau ^{2}}+%
\frac{1}{n}\left( \frac{\mu _{0}}{\tau ^{2}}\right)
^{2}\sigma^{2}+O_{p}\left( n^{-3/2}\right)
\end{equation*}
where $2\overline{y}\frac{\mu _{0}}{\tau ^{2}}$ is the order of $n^{-1/2}$,
and $-\overline{y}^{2}\frac{1}{\tau ^{2}}+\frac{1}{n}\left( \frac{\mu _{0}}{%
\tau ^{2}}\right) ^{2}\sigma^{2}$ has the order $n^{-1}$.
\end{remark}
Since $\mathbf{T}({\mbox{\boldmath${\mathbf{y}}$}},{%
\mbox{\boldmath${
\theta_0}$}})$ is calculated by using the MCMC output, it is important to
assess the numerical standard error for measuring the magnitude of
simulation error.
\begin{corollary}
\label{corol1}
Given the posterior draws $\{{\mbox{\boldmath${\vartheta}$}}^{\left( j\right) },j=1,2,\cdots ,J\}$, the numerical standard error (NSE) of the statistic $\mathbf{T}(\mathbf{y},\theta _{0})$ is,
\begin{equation*}
NSE\left( \widehat{\mathbf{T}}\left( \mathbf{y},{\mbox{\boldmath${\theta}$}}
_{0}\right) \right) = \sqrt{\frac{\partial \widehat{\mathbf{T}}\left( \mathbf{y},{\ %
\mbox{\boldmath${\theta}$}}_{0}\right) }{\partial \widehat{\mathbf{h}}}
Var\left( \widehat{{\mbox{\boldmath${h}$}}}\right) \left( \frac{\partial
\widehat{\mathbf{T}}\left( \mathbf{y},{\mbox{\boldmath${\theta}$}}
_{0}\right) }{\partial \widehat{{\mbox{\boldmath${h}$}}}}\right) ^{\prime }},
\end{equation*}%
where $\frac{\partial \widehat{\mathbf{T}}\left( \mathbf{y},{%
\mbox{\boldmath${
\theta}$}}_{0}\right) }{\partial \widehat{{\mbox{\boldmath${h}$}}}} =
-vec(\mbf{A}^\prime)^{\prime}\left(\widehat{{\mbox{\boldmath${H}$}}}%
^{\prime-1}\otimes\widehat{{\mbox{\boldmath${H}$}}}^{-1}\right)\frac{
\partial \widehat{{\mbox{\boldmath${H}$}}}}{\partial \widehat{{\ %
\mbox{\boldmath${h}$}}}}$, $\widehat{{
\mbox{\boldmath${H}$}}}=\frac{1}{J}\sum_{j=1}^{J}\left( {\ %
\mbox{\boldmath${\theta}$}}^{\left( j\right) }-{%
\mbox{\boldmath${\bar{
\theta}}$}}\right) \left( {\mbox{\boldmath${\theta}$}}^{\left(j\right) }-{\ %
\mbox{\boldmath${\bar{\theta}}$}}\right) ^{\prime }$,  $\mathbf{A} = \left({\mbox{\boldmath${\bar\theta-\theta_{0}}$}}\right)\left({\mbox{\boldmath${\bar\theta-\theta_{0}}$}}%
\right)^\prime$, $\widehat{\mathbf{h}}=vech\left(\widehat{{\mbox{\boldmath${H}$}}}\right)$, $\frac{ \partial \widehat{{\mbox{\boldmath${H}$}}}}{\partial \widehat{{\ %
\mbox{\boldmath${h}$}}}} = \left( \frac{\partial vec(\widehat{{%
\mbox{\boldmath${H}$}}})}{ \partial \widehat{{\mbox{\boldmath${h}$}}}}%
\right)$, $Var\left({\widehat{\mbf{h}}}\right)$ is the NSE of $\widehat{\mbf{h}}$.
\end{corollary}
The Corollary \ref{corol1} shows us how to compute the numerical standard error of the proposed statistic. For the NSE of $\widehat{\mbf{h}}$, $Var\left({\widehat{\mbf{h}}}\right)$, following Newey and West (1987), a consistent estimator can be given by
\begin{equation*}
Var(\widehat{\mathbf{h}})=\frac{1}{J}\left[ \Omega _{0}+\sum_{k=1}^{q}\left(
1-\frac{k}{q+1}\right) \left( \Omega _{k}+\Omega _{k}^{\prime }\right) %
\right] ,
\end{equation*}%
where
\begin{equation*}
\Omega _{k}=J^{-1}\sum_{j=k+1}^{J}\left( \mathbf{h}^{\left(j\right) }-
\widehat{\mathbf{h}}\right) \left( \mathbf{h}^{\left(j\right) }-\widehat{
\mathbf{h}}\right) ^{\prime }.
\end{equation*}
and the value of $q$ is always equal to 10.
\section{The Extension of the Test}
In this section, we extend the point-null hypothesis aforementioned
into the following problem,
\begin{equation*}
\begin{cases}
H_{0}: & R\mbf{\vartheta}_{0}=\mbf{r}\\
H_{1}: & R\mbf{\vartheta}_{0}\neq\mbf{r}
\end{cases},
\end{equation*}
where $R$ is a $m\times\left(d+q\right)$ matrix, $\mbf{r}\in\mathbb{R}^{m}$.
This hypothesis problem is much more general than the previous one.
On the other hand, it can help us to study the relationship among
parameters. Further, for such problems, it is hard to use the Bayes factor. Hence, the extension here is meaningful.

For such problem, the frequentist Wald statistic is
\begin{equation*}
\text{\textbf{Wald}}=\left(R\widehat{\mbf{\vartheta}}-\mbf{r}\right)^{\prime}\left[R\left(-\ddot{\mathcal{L}}_{n}^{-1}\left(\widehat{\mbf{\vartheta}}\right)\right)R^{\prime}\right]^{-1}\left(R\widehat{\mbf{\vartheta}}-\mbf{r}\right),
\end{equation*}
where $\widehat{\mbf{\vartheta}}$ is the MLE estimator of $\mbf{\vartheta}$.

According to the decision theory, we define the net loss function
for such problem as
\begin{equation*}
\Delta\mathcal{L}\left(H_{0},\mbf{\vartheta}\right)=\left(R\mbf{\vartheta}-\mbf{r}\right)^{\prime}\left[RV\left(\mbf{\bar{\vartheta}}\right)R^{\prime}\right]^{-1}\left(R\mathbf{\mbf{\vartheta}}-\mbf{r}\right),
\end{equation*}
where $\bar{\mbf{\vartheta}}$ is the posterior mean of $\mbf{\vartheta}$,
$V\left(\bar{\mbf{\vartheta}}\right)=E\left[\left.\left(\mbf{\vartheta}-\bar{\mbf{\vartheta}}\right)\left(\mbf{\vartheta}-\bar{\mbf{\vartheta}}\right)^{\prime}\right|\mbf{y},H_{1}\right]$.
Then the statistic is defined as
\begin{equation*}
\mbf{T}\left(\mbf{y},\mbf{r}\right)=\int\Delta\mathcal{L}\left(H_{0},\mbf{\vartheta}\right)d\mbf{\vartheta}=\int\left(R\mbf{\vartheta}-\mbf{r}\right)^{\prime}\left[RV\left(\mbf{\bar{\vartheta}}\right)R^{\prime}\right]^{-1}\left(R\mathbf{\mbf{\vartheta}}-\mbf{r}\right)d\mbf{\vartheta}.
\end{equation*}

\begin{theorem}
\label{thm2}Under Assumptions \textasciitilde{}\textasciitilde{}\textasciitilde{}\textasciitilde{},
when the likelihood information dominates the prior information, under
the null hypothesis,
\begin{equation*}
\mbf{T}\left(\mbf{y},\mbf{r}\right)-m=\text{\textbf{Wald}}+o_{p}\text{\ensuremath{\left(1\right)} \ensuremath{\overset{d}{\rightarrow}}}\chi^{2}\left(m\right).
\end{equation*}
\end{theorem}
Similarly, for the statistic, the numerical standard error can be computed in the following corollary.
\begin{corollary}
\label{corol2}
Given the posterior draws $\{{\mbox{\boldmath${\vartheta}$}}^{\left( j\right) },j=1,2,\cdots ,J\}$, the numerical standard error (NSE) of the statistic $\mathbf{T}(\mathbf{y},\mbf{r})$ is,
\begin{equation*}
NSE\left( \widehat{\mathbf{T}}\left( \mathbf{y},\mbf{r}\right) \right) = \sqrt{\frac{\partial \widehat{\mathbf{T}}\left( \mathbf{y},
\mbf{r}\right) }{\partial \widehat{\mathbf{h}}}
Var\left( \widehat{{\mbox{\boldmath${h}$}}}\right) \left( \frac{\partial
\widehat{\mathbf{T}}\left( \mathbf{y},\mbf{r}\right) }{\partial \widehat{{\mbox{\boldmath${h}$}}}}\right) ^{\prime }},
\end{equation*}%
where $\frac{\partial \widehat{\mathbf{T}}\left( \mathbf{y},
\mbf{r}\right) }{\partial \widehat{{\mbox{\boldmath${h}$}}}} =
-vec(\mbf{A}^\prime)^{\prime}\left[\left(R\widehat{\mbf{H}}R^{\prime} \right)^{-1}\otimes\left(R\widehat{\mbf{H}}R^{\prime} \right)^{-1}\right]\left(R\otimes R\right)\frac{
\partial \widehat{{\mbox{\boldmath${H}$}}}}{\partial \widehat{{\ %
\mbox{\boldmath${h}$}}}}$, $\widehat{{
\mbox{\boldmath${H}$}}}=\frac{1}{J}\sum_{j=1}^{J}\left( \mbf{\vartheta}^{\left( j\right) }-\bar{\mbf{\vartheta}}\right) \left( \mbf{\vartheta}^{\left( j\right) }-\bar{\mbf{\vartheta}}\right) ^{\prime }$,  $\mathbf{A} = \left(R{\mbox{\boldmath${\bar\vartheta} - \mbf{r}$}}\right)\left(R{\mbox{\boldmath${\bar\vartheta}$}} - \mbf{r}\right)^\prime$, $\widehat{\mathbf{h}}=vech\left(\widehat{{\mbox{\boldmath${H}$}}}\right)$, $\frac{ \partial \widehat{{\mbox{\boldmath${H}$}}}}{\partial \widehat{{\ %
\mbox{\boldmath${h}$}}}} = \left( \frac{\partial vec(\widehat{{%
\mbox{\boldmath${H}$}}})}{ \partial \widehat{{\mbox{\boldmath${h}$}}}}%
\right)$, $Var\left({\widehat{\mbf{h}}}\right)$ is the NSE of $\widehat{\mbf{h}}$.
\end{corollary}
For the NSE of $\widehat{\mbf{h}}$, $Var\left({\widehat{\mbf{h}}}\right)$, we can still follow the way proposed by Newey and West (1987) to evaluated.
\section{Simulation Studies}

In this section, we do two simulation studies to check the empirical size and power of the proposed test statistic. The first example is a simple simulation examination based on linear regression model where the our proposed test statistic has analytical expression. We compare the size and power of the new statistics with the Wald statistic. In the second example, we use the stochastic volatility model with leverage effect, where Wald statistic can not be used, to study the size and power of our statistic.

\subsection{The empirical power and size of $\mbf{T}\left(\mbf{y},{\mbf{\beta}}_{0}\right)$ and Wald statistic for linear regression model}
In this subsection, we use the simple linear regression model to examine the empirical power and size of the proposed test statistic. The model we use is
\begin{equation*}
y_{i}=\mbf{x}_{i}^{\prime}\mbf{\beta}+\epsilon_{i},\epsilon_{i}\sim N\left(0,\sigma^{2}\right),i=1,\dots,n.
\end{equation*}
with $\mbf{x}_{i1}=1$. Let $\mbf{X}=\left(\mbf{x}_{1}^{\prime},\dots,\mbf{x}_{N}^{\prime}\right)^{\prime}$,
then we can rewrite the model in matrix form,
\begin{equation*}
\mbf{y}=\mbf{X}\mbf{\beta}+\mbf{\epsilon},
\end{equation*}
where $\mbf{y}=\left(y_{1},\dots,y_{n}\right)^{\prime}$, $\mbf{\epsilon}=\left(\epsilon_{1},\dots,\epsilon_{n}\right)^{\prime}$.

We are interested in the subvector of $\mbf{\beta}$, $\breve{\mbf{\beta}}$,
then $\mbf{\beta}=\left(\breve{\mbf{\beta}}^{\prime},\tilde{\mbf{\beta}}^{\prime}\right)^{\prime}$.
Here we want to test $H_{0}:\breve{\mbf{\beta}}=\breve{\mbf{\beta}}_{0}$
against $H_{1}:\breve{\mbf{\beta}}\neq\breve{\mbf{\beta}}_{0}$ and $H_{0}:R\mbf{\beta}={\mbf{r}}$
against $H_{1}:R\mbf{\beta}\neq{\mbf{r}}$.
Assume that the prior distribution for $\mbf{\beta}$ and $\sigma^{2}$
are normal and inverse gamma, respectively,
\begin{equation*}
\mbf{\beta}|\sigma^{2}\sim N\left(\mu_{0},\sigma^{2}V_{0}\right),\sigma^{2}\sim IG\left(a,b\right),
\end{equation*}
where $\mu_{0}$, $V_{0}$ and $a$, $b$ are hyperparameters.

The proposed statistic $\mbf{T}\left(\mbf{y},{\mbf{\beta}}_{0}\right)$ for the first problem is
\begin{equation*}
\mbf{T}\left(\mbf{y},\breve{\mbf{\beta}}_{0}\right)=p+\frac{v-2}{2s}\left(\bar{\breve{\mbf{\beta}}}_{H_{1}}-\breve{\mbf{\beta}}_{0}\right)^{\prime}\breve{V}^{*-1}\left(\bar{\breve{\mbf{\beta}}}_{H_{1}}-\breve{\mbf{\beta}}_{0}\right),
\end{equation*}
where $v=2a+n$, $s=b+\frac{1}{2}\left(\mu_{0}^{\prime}V_{0}^{-1}\mu_{0}+\mbf{y}^{\prime}\mbf{y}-\mu^{*\prime}V^{*-1}\mu^{*}\right)$,
$V^{*}=\left(V_{0}^{-1}+\mbf{X}^{\prime}\mbf{X}\right)^{-1}$,$\mu^{*}=V^{*}\left(V_{0}^{-1}\tilde{\mu}+\mbf{X}^{\prime}\mbf{y}\right)$
and $\breve{V}^{*}$ the submatrix of $V^{*}$ corresponding to $\breve{\mbf{\beta}}$.
$p$ is the dimension of $\breve{\mbf{\beta}}$ and $\bar{\breve{\mbf{\beta}}}_{H_{1}}$
is the posterior mean of $\breve{\mbf{\beta}}$ under $H_{1}$. The details is given in the Appendix \ref{smlexpl1}.

For the second hypothesis problem, it can be readily derived that
the statistic is
\begin{equation*}
\mbf{T}\left(\mbf{y},\mbf{r}\right)=m+\frac{v-2}{2s}\left(R\bar{\mbf{\beta}}_{H_{1}}-\mbf{r}\right)^{\prime}\left(RV^{*}R^{\prime}\right)^{-1}\left(R\bar{\mbf{\beta}}_{H_{1}}-\mbf{r}\right).
\end{equation*}

For simplicity, we consider the case in which $\mbf{\beta}=\left(\beta_{1},\beta_{2}, \beta_{3},\beta_{4}\right)$, $\mbf{x}_{i}=\left(x_{i1},x_{i2},x_{i3},x_{i4}\right)^{\prime}$, where $x_{i1}=1$, $x_{i1},x_{i2},x_{i3},x_{i4}\sim N\left(0,1\right)$. In order to compare the empirical power and size between the new statistics and Wald statistic, the parameter values we use to simulate data are designed as $\sigma^{2}=0.01, \beta_{1} = 0.3, \beta_{2} = 0.2, \beta_{3} = 0.1\gamma, \beta_{4} = 0.5\gamma$ for $\gamma = 0, 0.1, 0.3, 0.5$. The replication number is 1000 and we consider the circumstances where the sample sizes are $n=50,100,150$, respectively.

In each replication, given the sample size, after the data simulation, we consider the hypothesis problem that whether $\beta_{2}=0$,$\beta_{3}=0$ , $\beta_{2}=\beta_{3}=0$ and $\beta_{2} + \beta_{3}=0$. In order to estimate the parameters, the prior we use is
\begin{equation*}
\tilde{\mu}=\left(0,0,0,0\right)^{\prime},\tilde{V}=1000I_{4},
\end{equation*}
\begin{equation*}
a=0.0001,b=0.0001,
\end{equation*}
where $I_{4}$ is the $4\times4$ identity matrix. For each scenarios,
we draw 5000 samples from the posterior distribution and then use
the posterior samples to obtain the posterior mean.

Given the credit level $95\%$, the ratios of the replications that reject the null hypothesis are computed and listed in Table $\ref{smlexpl1table1}$ in different scenarios. From the table, on one hand, the empirical size for the new statistic is quite good and almost the same as Wald statistic. For all the hypothesis problems, the size is approaching $5\%$ as the sample size increase. On the other hand, the empirical power performs well similar to the Wald statitic. As the $\gamma$ becomes larger, which implies that the values of parameters are further away from zero, and the sample size increase, the empirical power of the new statistic goes to $100\%$. All in all, the empirical power and size of the new statistic are very good and almost the same as the those of Wald statistic.
\begin{table}[tbph]
\caption{The empirical sizes and powers for linear model in different scenarios}
\label{smlexpl1table1}
\begin{center}
\footnotesize
\begin{tabular}{c|c|cc|cc|cc|cc}
\hline
\multicolumn{1}{c}{} &  & \multicolumn{2}{c|}{Empirical Size} & \multicolumn{6}{c}{Empirical Power}\\
\cline{3-10}
\multicolumn{1}{c}{} &  & \multicolumn{2}{c|}{$\gamma=0$} & \multicolumn{2}{c|}{$\gamma=0.1$} & \multicolumn{2}{c|}{$\gamma=0.3$} & \multicolumn{2}{c}{$\gamma=0.5$}\\
\hline
 & Null Hypothsis & $\boldsymbol{T}\left(\boldsymbol{y},\breve{\boldsymbol{\beta}}_{0}\right)$ & Wald  & $\boldsymbol{T}\left(\boldsymbol{y},\breve{\boldsymbol{\beta}}_{0}\right)$ & Wald  & $\boldsymbol{T}\left(\boldsymbol{y},\breve{\boldsymbol{\beta}}_{0}\right)$ & Wald  & $\boldsymbol{T}\left(\boldsymbol{y},\breve{\boldsymbol{\beta}}_{0}\right)$ & Wald \\
\hline
\multirow{4}{*}{$n=50$} & $\beta_{3}=0$ & $4.50\%$ & $5.10\%$ & $10.40\%$ & $11.00\%$ & $55.80\%$ & $57.30\%$ & $92.00\%$ & $92.20\%$\\
 & $\beta_{4}=0$ & $6.50\%$ & $7.10\%$ & $92.00\%$ & $92.5\%$ & $100\%$ & $100\%$ & $100\%$ & $100\%$\\
 & $\beta_{3}=\beta_{4}=0$ & $6.60\%$ & $7.50\%$ & $88.80\%$ & $89.70\%$ & $100\%$ & $100\%$ & $100\%$ & $100\%$\\
 & $\beta_{3}+\beta_{4}=0$ & $6.20\%$ & $6.70\%$ & $83.30\%$ & $84.00\%$ & $100\%$ & $100\%$ & $100\%$ & $100\%$\\
\hline
\multirow{4}{*}{$n=100$} & $\beta_{3}=0$ & $5.50\%$ & $5.80\%$ & $20.20\%$ & $20.40\%$ & $82.00
$ & $82.80\%$ & $99.90\%$ & $100\%$\\
 & $\beta_{4}=0$ & $4.60\%$ & $5.00\%$ & $99.70\%$ & $99.70\%$ & $100\%$ & $100\%$ & $100\%$ & $100\%$\\
 & $\beta_{3}=\beta_{4}=0$ & $5.70\%$ & $6.00\%$ & $99.50\%$ & $99.50\%$ & $100\%$ & $100\%$ & $100\%$ & $100\%$\\
 & $\beta_{3}+\beta_{4}=0$ & $6.00\%$ & $6.20\%$ & $98.60\%$ & $98.60\%$ & $100\%$ & $100\%$ & $100\%$ & $100\%$\\
\hline
\multirow{4}{*}{$n=150$} & $\beta_{3}=0$ & $5.30\%$ & $5.40\%$ & $24.40
$ & $24.60\%$ & $95.90\%$ & $95.90\%$ & $100\%$ & $100\%$\\
 & $\beta_{4}=0$ & $5.20\%$ & $5.30\%$ & $100\%$ & $100\%$ & $100\%$ & $100\%$ & $100\%$ & $100\%$\\
 & $\beta_{3}=\beta_{4}=0$ & $5.40\%$ & $5.60\%$ & $100\%$ & $100\%$ & $100\%$ & $100\%$ & $100\%$ & $100\%$\\
 & $\beta_{3}+\beta_{4}=0$ & $4.20\%$ & $4.20\%$ & $99.80\%$ & $99.80\%$ & $100\%$ & $100\%$ & $100\%$ & $100\%$\\
\hline
\end{tabular}
\end{center}
\end{table}

\subsection{The power and size of $\mbf{T}\left(\mbf{y},{\mbf{\beta}}_{0}\right)$  for leverage stochastic volatility model}
In this subsection, we examine the empirical power and size of the new statistic in stochastic volatility model with leverage effect (LSV). It is a type of latent variable models, for which the usual frequentist hypothesis tests such as Wald test can not be applied. But as we emphasize above, our new statistic can be readily used for such models. The model we study is defined as follows,
\begin{equation*}
\begin{cases}
r_{t}=\exp\left(\frac{h_{t}}{2}\right)\epsilon_{t},\\
h_{t+1}=\mu+\phi\left(h_{t}-\mu\right)+\sigma\varepsilon_{t+1},
\end{cases}
\end{equation*}
with
\begin{equation*}
\left(\begin{array}{c}
\epsilon_{t}\\
\varepsilon_{t+1}
\end{array}\right)\sim N\left(\left(\begin{array}{c}
0\\
0
\end{array}\right),\left(\begin{array}{cc}
1 & \rho\\
\rho & 1
\end{array}\right)\right),
\end{equation*}
where $r_{t}$ is the data observed, $h_{t}$ the latent volatility at period $t$. $\rho$ is the leverage effect. $\mu$, $\phi$ and $\sigma$ are the parameters we need to estimate. In order to examine the empirical power and size of the new hypoethesis testing, we use several sets of parameter values to simulate the model. We consider $\mu = -10, \phi = 0.97, \sigma^2 = 0.025$ and besides, $\rho = 0,-0.1,-0.2,-0.4$, respectively. The number of replications is 500 with sample size $T=1000,1500,2000$, respectively.

Then given the sample size $T$, we would like to test whether $\rho=0$
or not. That is,
\begin{equation*}
H_{0}:\rho=0,\text{ }H_{1}:\rho\neq0.
\end{equation*}
The priors we use to estimate the model in each case are listed
in the following,
\begin{equation*}
\mu\sim N\left(0,100\right),\phi\sim Beta\left(1,1\right),\sigma^{-2}\sim\Gamma\left(0.001,0.001\right),\rho\sim U\left(-1,1\right).
\end{equation*}
We use the R2OpenBUGS package to estimate the parameters. We draw 30,000 samples and the first 10,000 is discarded. The remaining 20,000 samples are used to compute the posterior means and statistic. Given the credit level $95\%$, the ratios of the replications that reject the null hypothesis are computed and listed in Table $\ref{smlexpl2table2}$ given different sample size.

From the Table $\ref{smlexpl2table2}$, on one hand, we can find that the empirical power of the new statistic performs well increasingly as the sample size increases. On the other hand, the empirical size also approaching $5.4\%$ as the sample size increases. To conclude, even for latent variable models, in which case usual methods are unavailable, our new statistic also possesses satisfactory power and size properties.

\begin{table}[tbph]
\caption{The rejection ratios of the new statistic for LSV model given credit
level $95\%$}
\label{smlexpl2table2}
\begin{center}
\begin{tabular}{c|c|ccc}
\hline
 & Empirical Size & \multicolumn{3}{c}{Empirical Power}\\
\hline
 & $\rho=0$ & $\rho=-0.1$ & $\rho=-0.2$ & $\rho=-0.4$\\
\hline
$T=1000$ & $7.80\%$ &  &  & $80.60\%$\\
$T=1500$ & $7.00\%$ &  &  & $93.60\%$\\
$T=2000$ & $5.40\%$ &  &  & $98.20\%$\\
\hline
\end{tabular}
\end{center}
\end{table}

\section{Empirical Illustrations}
In this section, we illustrate the proposed test statistic using two popular examples in economics and finance. The first example is a Multi-level Probit model. In this example, the observed data likelihood is available in closed-form, facilitating the comparison of the BF, the statistic in LLY and our proposed test. The second example is a stochastic volatility model with leverage effect, which is a typical case of latent variable model, where the volatility is latent.

\subsection{Examining the marginal effects on Probit model}

Li (2006) proposed a Bayesian method to estimate a simultaneous equation model. In her model, the first part is ordered Probit model. The second part is a two-limit censored regression.  She tried to examine the effect of high school education on income and unemployment period. Following her experiment, we use the same model and the same data set to implement our new test .

Let $z_{hi}$ denote the high school grade completed by individual $i$, and $%
y_{hi}$ denote the latent outcome corresponding to $z_{hi}$, where $h$
labels the schooling outcome, $z_{hi}=1$ if individual $i$ dropped out of
high school after completing the ninth grade, $z_{hi}=2$ if he dropped out
after completing the tenth grade, $z_{hi}=3$ if he dropped out after
completing the eleventh grade, and $z_{hi}=4$ if he completed high school.

\begin{equation}
\begin{cases}
y_{hi}={\mbox{\boldmath${\beta}$}}_{h}^{\prime}{\mbox{\boldmath${x}$}}%
_{hi}+\epsilon_{hi,} & \epsilon_{hi}\sim N\left(0,\sigma{}%
_{h}^{2}\right),\gamma_{z_{hi}}<y_{hi}<\gamma_{z_{hi}+1} \\
\gamma_{1}=-\infty,\gamma_{2}=0, & \gamma_{2}<\gamma_{3}<\gamma_{4},%
\gamma_{4}=1,\gamma_{5}=\infty%
\end{cases}%
,
\end{equation}
for $i=1,\dots,N,$ where ${\mbox{\boldmath${x}$}}_{hi}$ is a $k_{h}\times1$
vector of individual-level variables, including base year congnitive test
score, parental income, parental education, number of siblings, gender,
race, county level employment growth rate between 1980 and 1982, a
fourth-order polynomial in age and a fourth-order polynomial in the time
eligible to drop out. $\epsilon_{hi}$ is the individual-level random term, $%
N\left(\mu,\sigma^{2}\right)$ the normal distribution with mean $\mu$ and
variance $\sigma^{2}$, $\sigma_{h}^{2}$ the variance of the
unobservables, $\left\{ \gamma_{j}\right\} _{j=1}^{5}$ are the cutoff
points, and $N$ is the total number of individuals.

Let $\omega_{ui}$ denote the proportion of time individual $i$ is
unemployed, and $y_{ui}$ the latent outcome corresponding to $\omega_{ui}$,
and $y_{ui}$ is limited as,
\begin{equation}
y_{ui}%
\begin{cases}
\leq0 & \omega_{ui}=0 \\
=\omega_{ui} & 0<\omega_{ui}<1 \\
\geq 1 & \omega_{ui}=1%
\end{cases}%
,
\end{equation}
then the censored regression is,
\begin{equation}
y_{ui}={\mbox{\boldmath${\beta}$}}_{u}^{\prime}{\mbox{\boldmath${x}$}}_{ui}+{%
\mbox{\boldmath${s}$}}_{i}^{\prime}{\mbox{\boldmath${\eta}$}}+
\epsilon_{ui},\epsilon_{ui}\sim N\left(0,\sigma{}_{u}^{2}\right)
\end{equation}
for $i=1,2,\dots,N$, where $x_{ui}$ is $k_{u}\times1$ vector of observed
variables, including base year cognitive test score, parental income,
parental education, number of siblings, gender, race, age and a dummy
variable indicating any post-secondary education. $\epsilon_{ui}$ is the
unobservable, and $\sigma_{u}^{2}$ is the variance.

In the model, ${\mbox{\boldmath${s}$}}_{i}$ is a $4\times1$ vector of dummy
variables indicating the high school grade completed by individual $i$. Let $%
{\mbox{\boldmath${s}$}}_{i}=\left(s_{i,1},s_{i,2},s_{i,3},s_{i,4}\right)^{%
\prime}$, then $s_{i,z_{hi}}=1$ and $s_{i,j}=0$, $j\neq z_{hi}$. ${%
\mbox{\boldmath${\eta}$}}$ indicates the $4\times1$ vector of school
coefficients of $s_{i}$, which is different from the model in Li (2006).

The random terms are correlated,
\begin{equation*}
\left(%
\begin{array}{c}
\epsilon_{hi} \\
\epsilon_{ui}%
\end{array}%
\right)\sim N\left(\left(%
\begin{array}{c}
0 \\
0%
\end{array}%
\right),\left(%
\begin{array}{cc}
\sigma_{h}^{2} & \sigma_{hu} \\
\sigma_{hu} & \sigma_{u}^{2}%
\end{array}%
\right)\right)=N\left(0_{2\times1},\Sigma\right).
\end{equation*}

In the paper, the author used Bayesian method to estimate the parameters.
The priors she used are listed in the following.

\begin{equation*}
{\mbox{\boldmath${\beta}$}}=\left({\mbox{\boldmath${\beta}$}}_{h}^{\prime},{%
\mbox{\boldmath${\beta}$}}_{u}^{\prime}\right)^{\prime} \sim N\left({%
\mbox{\boldmath${\beta}$}}_{0},V_{\beta}\right),\ \Sigma\sim
IW\left(\rho,\rho R\right),
\end{equation*}
\begin{equation*}
{\mbox{\boldmath${\eta}$}}\sim N\left({\mbox{\boldmath${\eta}$}}%
_{0},V_{\eta}\right),\ \gamma_{3}\sim Beta\left(u_{1},u_{2}\right),
\end{equation*}
where ${\mbox{\boldmath${\beta}$}}_{0}=0_{k\times1}$, $k=k_{h}+k_{u}$, $%
V_{\beta}=1000I_{k}$, $IW\left(\rho,\rho R\right)$ denotes the inverted
Wishart distribution with degrees of freedom parameter $\rho$ and scale
parameter $R$, $\rho=6$, $R=I_{2}$, $\mbf{\eta}_{0}=0_{4\times1}$,$V_{\eta} = I_{4}$ , $u_{1}=u_{2}=1$, $Beta\left(\alpha,\delta\right)$ denotes the Beta distribution.

The estimation is almost the same as the the Gibbs method proposed by
Li (2006). We run the MCMC for 20,000 times. After dropping the first 4000
samples and convergence checking, we treat the left 16,000 as the effective
draws. The posterior means and the posterior standard errors are reported in
Table $\ref{empiricalexpl1table1}$.

\begin{table}[tbp]
\caption{The Posterior Means and Standard Errors of Parameters(without the
dummy variables)}
\label{empiricalexpl1table1}
\begin{center}
\begin{tabular}{clrrl}
\hline
& \multirow{2}{*}{} & $E\left(\cdot|Data\right)$ & $SE\left(\cdot|Data\right)
$ &  \\ \cline{3-4}
&  & \multicolumn{2}{l}{High school completion $y_{h}$} &  \\ \cline{2-4}
& Constant & 0.9474 & 0.2119 &  \\
& Parental income & 0.0110 & 0.0262 &  \\
& Base year cognitive test & 0.4413 & 0.0370 &  \\
& Father\textquoteright s education & 0.0456 & 0.0131 &  \\
& Mother\textquoteright s education & 0.0627 & 0.0159 &  \\
& Number of siblings & -0.0370 & 0.0153 &  \\
& Female & -0.0694 & 0.0534 &  \\
& Minority & 0.3840 & 0.0664 &  \\
& County employment growth & -0.0132 & 0.0047 &  \\
& Age & -0.4150 & 0.0853 &  \\
& $\mbox{Age}^{2}$ & -0.1887 & 0.0766 &  \\
& $\mbox{Age}^{3}$ & -0.0333 & 0.0468 &  \\
& $\mbox{Age}^{4}$ & 0.0311 & 0.0148 &  \\
& Time eligible to drop out & 0.0932 & 0.0696 &  \\
& $\mbox{Time}^{2}$ & 0.0905 & 0.0473 &  \\
& $\mbox{Time}^{3}$ & -0.0090 & 0.0106 &  \\
& $\mbox{Time}^{4}$ & -0.0094 & 0.0053 &  \\
&  & \multicolumn{2}{l}{Proportion of time unemployed $\omega_{u}$} &  \\
\cline{2-4}
& Parental income & -0.0275 & 0.0056 &  \\
& Base year cognitive test & -0.0392 & 0.0071 &  \\
& Father\textquoteright s education & -0.0020 & 0.0025 &  \\
& Mother\textquoteright s education & -0.0043 & 0.0030 &  \\
& Number of siblings & 0.0049 & 0.0034 &  \\
& Post-secondary education & -0.0113 & 0.0138 &  \\
& Female & 0.0621 & 0.0112 &  \\
& Minority & 0.0826 & 0.0131 &  \\
& Age & -0.0058 & 0.0126 &  \\
& Completing ninth grade(${\mbox{\boldmath${\eta}$}}_{1}$) & 0.1925 & 0.0705
&  \\
& Completing tenth grade(${\mbox{\boldmath${\eta}$}}_{2}$) & 0.1211 & 0.0530
&  \\
& Completing eleventh grade(${\mbox{\boldmath${\eta}$}}_{3}$) & 0.1187 &
0.0492 &  \\
& Completing high school(${\mbox{\boldmath${\eta}$}}_{4}$) & 0.0083 & 0.0416
&  \\
&  & \multicolumn{2}{l}{Civariance matrix $\Sigma$} &  \\ \cline{2-4}
& $\sigma_{h}^{2}$ & 0.9450 & 0.0914 &  \\
& $\sigma_{u}^{2}$ & 0.1215 & 0.0039 &  \\
& $\sigma_{hu}$ & -0.0099 & 0.0191 &  \\
&  & \multicolumn{2}{l}{Cutoff point} &  \\ \cline{2-4}
& $\gamma_{3}$ & 0.6684 & 0.0220 &  \\ \hline
\end{tabular}
\end{center}
\end{table}

In this example, we try to examine whether the marginal effects of father's
education $\left(\beta_{4}\right)$ and mother's education $\left(\beta_{5}%
\right)$ on the completion of high school can be ignored or not. Since the ${%
\mbox{\boldmath${T}$}}\left(Data,{\mbox{\boldmath${\theta}$}}_{0}\right)$
does not have analytical expression, according to the Appendix \ref{empexpl1}, we use the
MCMC output to approximate the statistic. Further, in order to compare the
statistic with the one in Li, Liu and Yu (2015) and the Bayes factor, we also
report the $\widehat{\log BF_{10}}$ and the $\widehat{{\mbox{\boldmath${T}$}}%
}_{LLY}\left(Data,{\mbox{\boldmath${\theta}$}}_{0}\right)$ in the Table \ref{empiricalexpl1table2}.
In this case, the log-likelihood has closed-form expression. Hence, the corresponding numerical standard error for each statistic is also reported in the Table $\ref{empiricalexpl1table2}$.

\begin{table}[tbp]
\caption{The proposed statistic $\protect\widehat{{\mbox{\boldmath${T}$}}}%
\left(Data,{\mbox{\boldmath${\theta}$}}_{0}\right)$,$\protect\widehat{{
\mbox{\boldmath${T}$}}}_{LLY}\left(Data,{\mbox{\boldmath${\theta}$}}%
_{0}\right)$, $\protect\widehat{\log BF_{10}}$, their computing time(in
seconds), and the numerical standard errors.}
\label{empiricalexpl1table2}
\begin{center}
\begin{tabular}{cccc}
\hline
& \multicolumn{3}{c}{$\beta_{3}=\beta_{4}=0$} \\ \cline{2-4}
& Value & NSE & Time \\ \hline
$\widehat{{\mbox{\boldmath${T}$}}}\left(Data,{\mbox{\boldmath${\theta}$}}%
_{0}\right)$ & 45.39 & 1.59 & 48311.31 \\
$\widehat{{\mbox{\boldmath${T}$}}}_{LLY}\left(Data,{\mbox{\boldmath${%
\theta}$}}_{0}\right)$ & 2502.00 & 89.57 & 87385.55 \\
$\widehat{\log BF_{10}}$ & 5.2019 & 1.03 & 341175.45 \\ \hline
\end{tabular}
\end{center}
\end{table}

The result we obtained in the Table $\ref{empiricalexpl1table2}$ strongly prove the
advantages of the proposed statistic. The 99.99 percentile of $%
\chi^{2}\left(2\right)$ is 18.42. Both the $\widehat{{\mbox{\boldmath${T}$}}}%
\left(Data,{\mbox{\boldmath${\theta}$}}_{0}\right)$ and $\widehat{{%
\mbox{\boldmath${T}$}}}_{LLY}\left(Data,{\mbox{\boldmath${\theta}$}}%
_{0}\right)$ are much larger than 18.42, which indicates that the null
hypothesis is rejected under the $99.99\%$ probability level. Those results
are consistent with the value of $\widehat{\log BF_{10}}$, which strongly
supports the alternative hypothesis. Those three statistics all tell us that
the marginal effect of the parents' education on the high school completion
is not negligible. Further, they all have small numerical standard error
compared with the corresponding values.

What's more from the table we can learn is that the proposed statistic takes
much less time than the other two statistics. It takes around as half as the
time $\widehat{{\mbox{\boldmath${T}$}}}_{LLY}\left(Data,{\mbox{\boldmath${%
\theta}$}}_{0}\right)$ used and one fifth of the time Bayes facor used.

\begin{remark}
From the example above, we can readily find that for the problem of high-dimensional parameters, the computation of the new statistic avoids the inversion of the large-scale information matrix, which is inevitable when we use Wald statistic. As we know, when the matrix is of large-scale, the information matrix may not be positive definite, therefore is singular. However, by using our new statistic, we only need compute the posterior covariance, which are the byproduct of the estimation procedure. Therefore, our statistic is superior to the Wald statistic for the hypothesis problems in a problem with many parameters.
\end{remark}

\subsection{Testing the leverage effect on the stochastic volatility models}

Stochastic volatility models are widely used in finance and economics. The
financial leverage effect  is very important and documented in many
financial literature, see Black (1976). Following Yu (2005), the leverage
effects SV model is defined as follows:

\begin{equation*}
\begin{cases}
r_{t}=\exp\left(\frac{h_{t}}{2}\right)\epsilon_{t} \\
h_{t+1}=\mu+\phi \left(h_{t} - \mu\right)+\sigma\varepsilon_{t+1}
\end{cases}%
\end{equation*}
with
\begin{equation*}
\left(%
\begin{array}{c}
\epsilon_{t} \\
\varepsilon_{t+1}%
\end{array}%
\right)\overset{i.i.d.}{\sim}N\left(\left(%
\begin{array}{c}
0 \\
0%
\end{array}%
\right),\left(%
\begin{array}{cc}
1 & \rho \\
\rho & 1%
\end{array}%
\right)\right),
\end{equation*}
and $h_{0}=\mu$, where $r_{t}$ is the return at time $t$, $h_{t}$ the return
volatility at period t. In this model, $\rho$ is the parameter indicating the
leverage effect. When $\rho<0$, there is a negative relationship between the
expected future volatility and the current return (Yu, 2005). In particular,
volatility tends to rise in response to bad news but fall in response to
good news (Black, 1976). Hence, we construct the hypothesis, $H_{0}:\rho=0$,
to test whether the leverage effect exists or not.

In this example, we used two cases to illustrate how to use the proposed
statistic. And further, the statistic is also compared with the statistic
proposed by LLY, ${\mbox{\boldmath${T}$}}_{LLY}\left({\mbox{\boldmath${y}$}},%
{\mbox{\boldmath${\theta}$}}_{0}\right)$ and Bayes factor. The derivation of
the computation is given in the Appendix \ref{empexpl1}.

In the first case, we use the data that consist of daily returns on
Pound/Dollar exchange rates from 01/10/81 to 28/06/85 with sample size 945. The series $r_{t}$ is
the daily mean-corrected returns. The R2OpenBUGS is used to estimate the
model with the following priors for each parameter:
\begin{equation*}
\mu\sim N\left(0,100\right),\phi\sim
Beta\left(1,1\right),\sigma^{-2}\sim\Gamma\left(0.001,0.001\right),\rho\sim
U\left(-1,1\right).
\end{equation*}

\begin{table}[tbph]
\caption{The posterior mean of parameter estimated in case 1}
\label{empiricalexpl2table1}
\begin{center}
\vspace{1em}
\begin{tabular}{ccccc}
\hline
& \multicolumn{2}{c}{$H_{1}$} & \multicolumn{2}{c}{$H_{0}$} \\ \hline
Parameter & Mean & SE & Mean & SE \\ \hline
$\mu$ & -0.5776 & 0.3487 & -0.6608 & 0.3164 \\
$\phi$ & 0.9849 & 0.0097 & 0.9793 & 0.0127 \\
$\rho$ & -0.0941 & 0.1507 & - & - \\
$\tau$ & 0.1553 & 0.0243 & 0.1618 & 0.0360 \\ \hline
\end{tabular}%
\end{center}
\end{table}

We draw 50,000 from the posterior distribution and discard the first 20,000 as build-in period. Then we store every 5th value of the remaining samples as effective observations. The estimation
results are reported in Table $\ref{empiricalexpl2table1}$.

We aim to test whether there is leverage effect or not, hence the hypothesis problem is:

\begin{equation*}
H_{0}:\rho=0,\,\,H_{1}:\rho\neq0.
\end{equation*}

\begin{table}[tbph]
\caption{The statistic $\protect\widehat{{\mbox{\boldmath${T}$}}}_{LLY}\left(%
{\ \mbox{\boldmath${y}$}},{\mbox{\boldmath${\theta}$}}_{0}\right)$, $\protect%
\widehat{{\ \mbox{\boldmath${T}$}}}\left({\mbox{\boldmath${y}$}},{
\mbox{\boldmath${\theta}$}}_{0}\right)$, $\protect\widehat{\log BF_{10}}$,
their computing time (in seconds), and the numerical standard errors of the
first two statistics in case 1.}
\label{empiricalexpl2table2}
\begin{center}
\vspace{1em}
\begin{tabular}{cccc}
\hline
& $\widehat{{\mbox{\boldmath${T}$}}}\left({\mbox{\boldmath${y}$}},{%
\mbox{\boldmath${\theta}$}}_{0}\right)$ & $\widehat{{\mbox{\boldmath${T}$}}}%
_{LLY}\left({\mbox{\boldmath${y}$}},{\mbox{\boldmath${\theta}$}}_{0}\right)$
& $\widehat{\log BF_{10}}$ \\ \hline
Value & 1.3893 & 0.2883 & -10.1235 \\
NSE & 0.0255 & 0.2028 & - \\
Time used(s) & 1765.3591 & 2313.4812 & 5465.6422 \\ \hline
\end{tabular}%
\end{center}
\end{table}

In Table $\ref{empiricalexpl2table2}$, we report the Bayes factor, the statistic in LLY, $%
\widehat{{\mbox{\boldmath${T}$}}}_{LLY}\left({\mbox{\boldmath${y}$}},{%
\mbox{\boldmath${\theta}$}}_{0}\right)$, and the proposed statistic, $%
\widehat{{\mbox{\boldmath${T}$}}}\left({\mbox{\boldmath${y}$}},{%
\mbox{\boldmath${\theta}$}}_{0}\right)$. The $\widehat{\log BF_{10}}$
strongly supports the null hypothesis, that is, there is not leverage
effect. Meanwhile, since ${\mbox{\boldmath${T}$}}_{LLY}\left({%
\mbox{\boldmath${y}$}},{\mbox{\boldmath${\theta}$}}_{0}\right)$ follows a $%
\chi^{2}\left(1\right)$ distribution, the value of this statistic with a
rather small NSE shows that it fails to reject the null hypothesis at the
95\% probability level. For the porposed statistic, ${\mbox{\boldmath${T}$}}%
\left({\mbox{\boldmath${y}$}},{\mbox{\boldmath${\theta}$}}_{0}\right)-1%
\overset{d}{\rightarrow}\chi^{2}\left(1\right)$, $\widehat{{%
\mbox{\boldmath${T}$}}}\left({\mbox{\boldmath${y}$}},{\mbox{\boldmath${%
\theta}$}}_{0}\right)-1$ is closed to $\widehat{{\mbox{\boldmath${T}$}}}%
_{LLY}\left({\mbox{\boldmath${y}$}},{\mbox{\boldmath${\theta}$}}_{0}\right)$
with smaller NSE. Thus it can not reject the null hypothesis under 95\%
probability level. Thus, the outcomes of all three statistics are consistent.

In the second case, the data we use is 1,822 daily returns of the Standard \& Poor (S\&P) 500 index,
covering the period between January 3, 2005 and March 28, 2012. We use the same priors and similar method to estimate the model. And the parameter estimated are
listed in Table $\ref{empiricalexpl2table3}$, which are quite different from the first case.

\begin{table}[tbph]
\caption{The posterior mean of parameter estimated in case 2}
\label{empiricalexpl2table3}
\begin{center}
\vspace{1em}
\begin{tabular}{ccccc}
\hline
& \multicolumn{2}{c}{$H_{1}$} & \multicolumn{2}{c}{$H_{0}$} \\ \hline
Parameter & Mean & SE & Mean & SE \\ \hline
$\mu$ & -10.8800 & 0.1751 & -11.2200 & 0.3349 \\
$\phi$ & 0.9804 & 0.0039 & 0.9897 & 0.0042 \\
$\rho$ & -0.7151 & 0.0422 & - & - \\
$\tau$ & 0.2057 & 0.0178 & 0.1705 & 0.0169 \\ \hline
\end{tabular}%
\end{center}
\end{table}

Again, the three statistics are reported in Table $\ref{empiricalexpl2table4}$. Contrary to
first case, all the statistics strongly support the null hypotheses, that is,
there is leverage effect in the data. For $\widehat{{\mbox{\boldmath${T}$}}}%
\left({\mbox{\boldmath${y}$}},{\mbox{\boldmath${\theta}$}}_{0}\right)$ and $%
\widehat{{\mbox{\boldmath${T}$}}}_{LLY}\left({\mbox{\boldmath${y}$}},{%
\mbox{\boldmath${\theta}$}}_{0}\right)$, they all reject the null hypothesis
under the 99\% probability level. At the same time, the $\widehat{%
\log BF_{10}}$ also strongly supports the alternative hypothesis. Therefore,
the results of all three statistics are consistent.

\begin{table}[tbph]
\caption{The statistic $\protect\widehat{{\mbox{\boldmath${T}$}}}_{LLY}\left(%
{\ \mbox{\boldmath${y}$}},{\mbox{\boldmath${\theta}$}}_{0}\right)$, $\protect%
\widehat{{\ \mbox{\boldmath${T}$}}}\left({\mbox{\boldmath${y}$}},{
\mbox{\boldmath${\theta}$}}_{0}\right)$, $\protect\widehat{\log BF_{10}}$,
their computing time (in seconds), and the numerical standard errors of the
first two statistics in case 2.}
\label{empiricalexpl2table3}
\begin{center}
\vspace{1em}
\begin{tabular}{cccc}
\hline
& $\widehat{{\mbox{\boldmath${T}$}}}\left({\mbox{\boldmath${y}$}},{%
\mbox{\boldmath${\theta}$}}_{0}\right)$ & $\widehat{{\mbox{\boldmath${T}$}}}%
_{LLY}\left({\mbox{\boldmath${y}$}},{\mbox{\boldmath${\theta}$}}_{0}\right)$
& $\widehat{\log BF_{10}}$ \\ \hline
Value & 287.7944 & 8.2419 & 51.9582 \\
NSE & 0.6915 & 0.6849 & - \\
Time used(s) & 5606.8825 & 6862.3672 & 13391.2791 \\ \hline
\end{tabular}%
\end{center}
\end{table}

\section{Conclusion}

In this paper, a new $\chi^2$-type Bayesian test statistic is proposed to
test a point null hypothesis for latent variable models. The new statistic
can be explained as Bayesian version of Wald test. Compared with existing
literature, the proposed test statistic has achieved several important
advantages, hence, can appeal many practical applications. First, for latent
variable models, it is only by-product of posterior outputs, hence, is very
easy to compute, not require additional computational efforts after the
model is estimated using MCMC techniques. Second, it is well-defined under
improper prior distributions and avoids Jeffrey-Lindley's paradox. Third,
it's asymptotic distribution is pivotal so that the threshold values can be
easily obtained from the asymptotic chi-squared distribution. Through Monte
Carlo studies, it can be shown that the test power is almost equivalent to
Wald test, but better than the test statistic by Li, Liu and Yu (2014). If
the observed likelihood doesn't have analytical form, Wald test statistic is
very difficult to be applied, but, the proposed test statistic is also easy
to implement.

\textbf{The Bayes factor and the Bayesian test statistic with Li, Liu and Yu
(2015) need to be computed out and make an empirical comparison }

\vskip 0.5cm

\section{Appendix }
\subsection{Proof of Theorem \protect\ref{thm1}}

%
%
%

According to LZY, we have
\begin{equation*}
E\left[\left({\mbox{\boldmath${\vartheta}$}}-\widehat{{\mbox{\boldmath${%
\vartheta}$}}}_m\right)|\mathbf{y}\right]=o_{p}\left(n^{-\frac{1}{2}}\right),
\end{equation*}
\begin{eqnarray*}
V\left({\mbox{\boldmath${\widehat{\vartheta}}$}}_m\right) & = & E\left[\left(%
{\mbox{\boldmath${\vartheta-\widehat{\vartheta}}$}}_m\right)\left({%
\mbox{\boldmath${\vartheta-\widehat{\vartheta}}$}}_m\right)^{\prime}|\mathbf{%
y}\right]= -\ddot{\mathcal{L}}^{-1}_n({\mbox{\boldmath${\widehat\vartheta}$}}%
_m)+o_{p}\left(n^{-1}\right),
\end{eqnarray*}
\begin{equation*}
\mbox{\boldmath${\widehat\vartheta}$}-{\mbox{\boldmath${\widehat\vartheta}$}}%
_m=o_p(n^{-\frac{1}{2}}).
\end{equation*}
Further, we can get
\begin{equation*}
-\ddot{\mathcal{L}}^{-1}_n({\mbox{\boldmath${\widehat\vartheta}$}}_m)=-\ddot{%
\mathcal{L}}^{-1}_n({\mbox{\boldmath${\widehat\vartheta}$}}%
)+o_{p}\left(n^{-1}\right),
\end{equation*}
by the Talyor expansion of $-\ddot{\mathcal{L}}^{-1}_n({\mbox{\boldmath${%
\widehat\vartheta}$}}_m)$ at ${\mbox{\boldmath${\widehat\vartheta}$}}$.
Hence we have
\begin{equation*}
V\left({\mbox{\boldmath${\bar{\vartheta}}$}}\right) = V\left({%
\mbox{\boldmath${\widehat{\vartheta}}$}}_m\right)+o_{p}\left(n^{-1}\right) =
-\ddot{\mathcal{L}}^{-1}_n({\mbox{\boldmath${\widehat\vartheta}$}}%
)+o_{p}\left(n^{-1}\right)
\end{equation*}
\begin{equation*}
 \bar{\mbf{\vartheta}} - \widehat{\mbf{\vartheta}} = o_p(n^{-\frac{1}{2}}),
\end{equation*}
Let $V_{\theta\theta}\left({\mbox{\boldmath${\bar{\vartheta}}$}}\right) $ is
the submatrix of $V\left({\mbox{\boldmath${\bar{\vartheta}}$}}\right)$
w.r.t. ${\mbox{\boldmath${\theta}$}}$ . Since ${\mbox{\boldmath${\widehat%
\theta}$}}_{m}-{\mbox{\boldmath${\theta}$}}_0=O_p(n^{-1/2})$, under the null
hypothesis, we can show that
\begin{eqnarray*}
& &~~~({\mbox{\boldmath${\widehat\theta}$}}-{\mbox{\boldmath${\theta}$}}%
_0)^{\prime}\left[\mathbf{V}_{\theta\theta}({\mbox{\boldmath${\bar%
\vartheta}$}})\right]^{-1}({\mbox{\boldmath${\widehat\theta}$}}-{%
\mbox{\boldmath${\theta}$}}_0) =({\mbox{\boldmath${\widehat\theta}$}}-{%
\mbox{\boldmath${\theta}$}}_0)^{\prime}\left[-\ddot{\mathcal{L}}%
_{n,\theta\theta}^{-1}({\mbox{\boldmath${\widehat\vartheta}$}})+o_p(n^{-1})%
\right]^{-1}({\mbox{\boldmath${\widehat\theta}$}}-{\mbox{\boldmath${%
\theta}$}}_0) \\
& &=\sqrt{n}({\mbox{\boldmath${\widehat\theta}$}}-{\mbox{\boldmath${\theta}$}%
}_0)^{\prime}\left[-n\ddot{\mathcal{L}}_{n,\theta\theta}^{-1}({%
\mbox{\boldmath${\widehat\vartheta}$}})+o_p(1)\right]^{-1}\sqrt{n}({%
\mbox{\boldmath${\widehat\theta}$}}-{\mbox{\boldmath${\theta}$}}_0) \\
& &=\sqrt{n}({\mbox{\boldmath${\widehat\theta}$}}-{\mbox{\boldmath${\theta}$}%
}_0)^{\prime}\left[-n\ddot{\mathcal{L}}_{n,\theta\theta}^{-1}({%
\mbox{\boldmath${\widehat\vartheta}$}})\right]^{-1}\sqrt{n}({%
\mbox{\boldmath${\widehat\theta}$}}-{\mbox{\boldmath${\theta}$}}_0)+o_p(1)%
\sqrt{n}({\mbox{\boldmath${\widehat\theta}$}}-{\mbox{\boldmath${\theta}$}}%
_0)^{\prime}\sqrt{n}({\mbox{\boldmath${\widehat\theta}$}}-{%
\mbox{\boldmath${\theta}$}}_0) \\
& &=\sqrt{n}({\mbox{\boldmath${\widehat\theta}$}}-{\mbox{\boldmath${\theta}$}%
}_0)^{\prime}\left[-n\ddot{\mathcal{L}}_{n,\theta\theta}^{-1}({%
\mbox{\boldmath${\widehat\vartheta}$}})\right]^{-1}\sqrt{n}({%
\mbox{\boldmath${\widehat\theta}$}}-{\mbox{\boldmath${\theta}$}}_0)+o_p(1)%
\sqrt{n}O_p(n^{-1/2})\sqrt{n}O_p(n^{-1/2}) \\
& &=\sqrt{n}({\mbox{\boldmath${\widehat\theta}$}}-{\mbox{\boldmath${\theta}$}%
}_0)^{\prime}\left[-n\ddot{\mathcal{L}}_{n,\theta\theta}^{-1}({%
\mbox{\boldmath${\widehat\vartheta}$}})\right]^{-1}\sqrt{n}({%
\mbox{\boldmath${\widehat\theta}$}}-{\mbox{\boldmath${\theta}$}}_0)+o_p(1) \\
&&=\mathbf{Wald}+o_p(1)\overset{d}{\rightarrow } \chi^2(p)
\end{eqnarray*}
It is noted that
\begin{eqnarray*}
& &V\left({\mbox{\boldmath${\vartheta}$}}_{0}\right)=E\left[({%
\mbox{\boldmath${\theta}$}}-{\mbox{\boldmath${\theta}$}}_0)({%
\mbox{\boldmath${\theta}$}}-{\mbox{\boldmath${\theta}$}}_0)^{\prime}|\mathbf{%
y}\right] \\
& = & E\left[\left({\mbox{\boldmath${\vartheta-\widehat{\vartheta}+%
\widehat{\vartheta}-\vartheta_{0}}$}}\right)\left({\mbox{\boldmath${%
\vartheta-\widehat{\vartheta}+\widehat{\vartheta}-\vartheta_{0}}$}}%
\right)^{\prime}|\mathbf{y}\right] \\
& = & E\left[\left({\mbox{\boldmath${\vartheta}$}}-{\mbox{\boldmath${%
\widehat{\vartheta}}$}}\right)\left({\mbox{\boldmath${\vartheta}$}}-{%
\mbox{\boldmath${\widehat{\vartheta}}$}}\right)^\prime|\mathbf{y}\right]+2E%
\left[\left({\mbox{\boldmath${\widehat{\vartheta}-\vartheta_{0}}$}}%
\right)\left({\mbox{\boldmath${\vartheta-\widehat{\vartheta}}$}}%
\right)^\prime|\mathbf{y}\right]+E\left[\left({\mbox{\boldmath${\widehat{%
\vartheta}-\vartheta_{0}}$}}\right)\left({\mbox{\boldmath${\widehat{%
\vartheta}-\vartheta_{0}}$}}\right)^{\prime}|\mathbf{y}\right] \\
& = & V\left({\mbox{\boldmath${\widehat\vartheta}$}}\right)+2\left({%
\mbox{\boldmath${\widehat{\vartheta}-\vartheta_{0}}$}}\right)\left({%
\mbox{\boldmath${\bar{\vartheta}-\widehat{\vartheta}}$}}\right)^{\prime}+%
\left({\mbox{\boldmath${\widehat{\vartheta}-\vartheta_{0}}$}}\right)\left({%
\mbox{\boldmath${\widehat{\vartheta}-\vartheta_{0}}$}}\right)^{\prime} \\
& = & V\left({\mbox{\boldmath${\widehat\vartheta}$}}\right)+2O_{p}\left(n^{-%
\frac{1}{2}}\right)o_{p}\left(n^{-\frac{1}{2}}\right)+\left({%
\mbox{\boldmath${\widehat{\vartheta}-\vartheta_{0}}$}}\right)\left({%
\mbox{\boldmath${\widehat{\vartheta}-\vartheta_{0}}$}}\right)^{\prime} \\
& = & V\left({\mbox{\boldmath${\widehat\vartheta}$}}\right)+\left({%
\mbox{\boldmath${\widehat{\vartheta}-\vartheta_{0}}$}}\right)\left({%
\mbox{\boldmath${\widehat{\vartheta}-\vartheta_{0}}$}}\right)^{\prime}+o_{p}%
\left(n^{-1}\right)
\end{eqnarray*}

In addition, we also can prove that
\begin{eqnarray*}
V\left({\mbox{\boldmath${\bar{\vartheta}}$}}\right) & = & E\left[\left({%
\mbox{\boldmath${\vartheta-\widehat{\vartheta}+\widehat{\vartheta}-\bar{%
\vartheta}}$}}\right)\left({\mbox{\boldmath${\vartheta-\widehat{%
\vartheta}+\widehat{\vartheta}-\bar{\vartheta}}$}}\right)^{\prime}|%
\mathbf{y}\right] \\
& = & E\left[\left({\mbox{\boldmath${\vartheta}$}}-{\mbox{\boldmath${%
\widehat{\vartheta}}$}}\right)\left({\mbox{\boldmath${\vartheta}$}}-{%
\mbox{\boldmath${\widehat{\vartheta}}$}}\right)^{\prime}|\mathbf{y}\right]+2E%
\left[\left({\mbox{\boldmath${\widehat{\vartheta}-\bar{\vartheta}}$}}%
\right)\left({\mbox{\boldmath${\vartheta-\widehat{\vartheta}}$}}%
\right)^{\prime}|\mathbf{y}\right]+E\left[\left({\mbox{\boldmath${\widehat{%
\vartheta}-\bar{\vartheta}}$}}\right)\left({\mbox{\boldmath${\widehat{%
\vartheta}-\bar{\vartheta}}$}}\right)\prime|\mathbf{y}\right] \\
& = & V\left({\mbox{\boldmath${\widehat\vartheta}$}}\right)+2\left({%
\mbox{\boldmath${\widehat{\vartheta}-\bar{\vartheta}}$}}\right)\left({%
\mbox{\boldmath${\bar{\vartheta}-\widehat{\vartheta}}$}}\right)^{\prime}+%
\left({\mbox{\boldmath${\widehat{\vartheta}-\bar{\vartheta}}$}}\right)\left({%
\mbox{\boldmath${\widehat{\vartheta}-\bar{\vartheta}}$}}\right)\prime \\
& = & V\left({\mbox{\boldmath${\widehat\vartheta}$}}\right)-\left({%
\mbox{\boldmath${\widehat{\vartheta}-\bar{\vartheta}}$}}\right)\left({%
\mbox{\boldmath${\widehat{\vartheta}}$}}- {\mbox{\boldmath${\bar{%
\vartheta}}$}}\right)^{\prime} \\
& = & V\left({\mbox{\boldmath${\widehat\vartheta}$}}\right)+o_{p}%
\left(n^{-1/2}\right)o_{p}\left(n^{-1/2}\right) \\
& = & V\left({\mbox{\boldmath${\widehat\vartheta}$}}\right)+o_{p}%
\left(n^{-1}\right)
\end{eqnarray*}

Hence, we can prove that
\begin{eqnarray*}
\mathbf{T}\left(\mathbf{y},{\mbox{\boldmath${\theta_{0}}$}}\right) & = &
\int\left({\mbox{\boldmath${\theta-\theta_{0}}$}}\right)\left[\mathbf{V}%
_{\theta\theta}({\mbox{\boldmath${\bar\vartheta}$}})\right]^{-1}\left({%
\mbox{\boldmath${\theta-\theta_{0}}$}}\right)^{\prime} d{\mbox{\boldmath${%
\vartheta}$}} \\
&=&\mathbf{tr}\left\{\left[\mathbf{V}_{\theta\theta}({\mbox{\boldmath${\bar%
\vartheta}$}})\right]^{-1}E\left[\left({\mbox{\boldmath${\theta-\theta_{0}}$}%
}\right)\left({\mbox{\boldmath${\theta-\theta_{0}}$}}\right)^\prime|\mathbf{y%
}\right]\right\} \\
&=&\mathbf{tr}\left\{\left[\left[\mathbf{V}_{\theta\theta}({%
\mbox{\boldmath${\widehat{\vartheta}}$}})\right]^{-1}+o_p(n)\right]E\left[%
\left({\mbox{\boldmath${\theta-\theta_{0}}$}}\right)\left({%
\mbox{\boldmath${\theta-\theta_{0}}$}}\right)^\prime|\mathbf{y}\right]%
\right\} \\
&=&\mathbf{tr}\left\{\left[\mathbf{V}_{\theta\theta}({\mbox{\boldmath${%
\widehat{\vartheta}}$}})\right]^{-1}E\left[\left({\mbox{\boldmath${\theta-%
\theta_{0}}$}}\right)\left({\mbox{\boldmath${\theta-\theta_{0}}$}}%
\right)^\prime|\mathbf{y}\right]\right\} +\mathbf{tr}\left\{o_p(n)E\left[%
\left({\mbox{\boldmath${\theta-\theta_{0}}$}}\right)\left({%
\mbox{\boldmath${\theta-\theta_{0}}$}}\right)^\prime|\mathbf{y}\right]%
\right\} \\
&=&\mathbf{tr}\left\{\left[\mathbf{V}_{\theta\theta}({\mbox{\boldmath${%
\widehat{\vartheta}}$}})\right]^{-1}E\left[\left({\mbox{\boldmath${\theta-%
\theta_{0}}$}}\right)\left({\mbox{\boldmath${\theta-\theta_{0}}$}}%
\right)^\prime|\mathbf{y}\right]\right\} +o_p(n)O_p(n^{-1}) \\
&=&\mathbf{tr}\left\{\left[\mathbf{V}_{\theta\theta}({\mbox{\boldmath${%
\widehat{\vartheta}}$}})\right]^{-1}E\left[\left({\mbox{\boldmath${\theta-%
\theta_{0}}$}}\right)\left({\mbox{\boldmath${\theta-\theta_{0}}$}}%
\right)^\prime|\mathbf{y}\right]\right\} +o_p(1) \\
&=&\mathbf{tr}\left\{\left[\mathbf{V}_{\theta\theta}({\mbox{\boldmath${%
\widehat{\vartheta}}$}})\right]^{-1}\left[\mathbf{V}_{\theta\theta}\left({%
\mbox{\boldmath${\widehat{\vartheta}}$}}\right)+\left({\mbox{\boldmath${%
\widehat{\theta}-\theta_{0}}$}}\right)\left({\mbox{\boldmath${\widehat{%
\theta}-\theta_{0}}$}}\right)^\prime+o_p(n^{-1})\right]\right\} +o_p(1) \\
&=&\mathbf{tr}\left\{\left[\mathbf{V}_{\theta\theta}({\mbox{\boldmath${%
\widehat{\vartheta}}$}})\right]^{-1}\left[\mathbf{V}_{\theta\theta}\left({%
\mbox{\boldmath${\widehat{\vartheta}}$}}\right)+\left({\mbox{\boldmath${%
\widehat{\theta}-\theta_{0}}$}}\right)\left({\mbox{\boldmath${\widehat{%
\theta}-\theta_{0}}$}}\right)^\prime\right]\right\} +\mathbf{tr}\left[\left[%
\mathbf{V}_{\theta\theta}({\mbox{\boldmath${\widehat{\vartheta}}$}})\right]%
^{-1}o_p(n^{-1})\right]+o_p(1) \\
&=&\mathbf{tr}\left\{\left[\mathbf{V}_{\theta\theta}({\mbox{\boldmath${%
\widehat{\vartheta}}$}})\right]^{-1}\left[\mathbf{V}_{\theta\theta}\left({%
\mbox{\boldmath${\widehat{\vartheta}}$}}\right)+\left({\mbox{\boldmath${%
\widehat{\theta}-\theta_{0}}$}}\right)\left({\mbox{\boldmath${\widehat{%
\theta}-\theta_{0}}$}}\right)^\prime\right]\right\} +o_p(1) \\
&=&p+\mathbf{tr}\left\{\left[\mathbf{V}_{\theta\theta}({\mbox{\boldmath${%
\widehat{\vartheta}}$}})\right]^{-1}\left[\left({\mbox{\boldmath${\widehat{%
\theta}-\theta_{0}}$}}\right)\left({\mbox{\boldmath${\widehat{\theta}-%
\theta_{0}}$}}\right)^\prime\right]\right\}+o_p(1) \\
&=&p+\left({\mbox{\boldmath${\widehat{\theta}-\theta_{0}}$}}\right)^\prime%
\left[\mathbf{V}_{\theta\theta}({\mbox{\boldmath${\widehat{\vartheta}}$}})%
\right]^{-1}\left({\mbox{\boldmath${\widehat{\theta}-\theta_{0}}$}}%
\right)+o_p(1) \\
&=&p+\mathbf{Wald}+o_p(1)
\end{eqnarray*}

Furthermore, it is easily showed that
\begin{eqnarray*}
\mathbf{T}\left(\mathbf{y},{\mbox{\boldmath${\theta_{0}}$}}\right)-p=\mathbf{%
Wald}+o_p(1)\overset{d}{\rightarrow } \chi^2(p)
\end{eqnarray*}

\subsection{Proof of Corollary \protect\ref{corol1}}
The statistic $\mathbf{T}({\mbox{\boldmath${\mathbf{y}}$}},%
{\mbox{\boldmath${            \theta_0}$}})$ can be rewritten as
\begin{eqnarray*}
\mathbf{T}\left(\mathbf{y},{\mbox{\boldmath${\theta_{0}}$}}\right) & = &
\int\left({\mbox{\boldmath${\theta-\theta_{0}}$}}\right)\left[\mathbf{V}%
_{\theta\theta}({\mbox{\boldmath${\bar\vartheta}$}})\right]^{-1}\left({%
\mbox{\boldmath${\theta-\theta_{0}}$}}\right)^{\prime} d{\mbox{\boldmath${%
\vartheta}$}} \\
&=&\int\left({\mbox{\boldmath${\theta-\theta_{0}}$}}\right)\left[\mathbf{V}%
_{\theta\theta}({\mbox{\boldmath${\bar\vartheta}$}})\right]^{-1}\left({%
\mbox{\boldmath${\theta-\theta_{0}}$}}\right)\prime d{\mbox{\boldmath${%
\vartheta}$}} \\
&=&\mathbf{tr}\left\{\left[\mathbf{V}_{\theta\theta}({\mbox{\boldmath${\bar%
\vartheta}$}})\right]^{-1}E\left[\left({\mbox{\boldmath${\theta-\theta_{0}}$}%
}\right)\left({\mbox{\boldmath${\theta-\theta_{0}}$}}\right)^\prime|\mathbf{y%
}\right]\right\} \\
&=&\mathbf{tr}\left\{\left[\mathbf{V}_{\theta\theta}({\mbox{\boldmath${\bar%
\vartheta}$}})\right]^{-1}\left[\mathbf{V}_{\theta\theta}\left({%
\mbox{\boldmath${\bar{\vartheta}}$}}\right)+\left({\mbox{\boldmath${\bar%
\theta-\theta_{0}}$}}\right)\left({\mbox{\boldmath${\bar\theta-\theta_{0}}$}}%
\right)^\prime\right]\right\} \\
&=&p + \mathbf{tr}\left\{\left({\mbox{\boldmath${\bar\theta-\theta_{0}}$}}%
\right)\left({\mbox{\boldmath${\bar\theta-\theta_{0}}$}}\right)^\prime\left[%
\int \left( {\mbox{\boldmath${\theta}$}}-{\ \mbox{\boldmath${\bar{\theta}}$}}%
\right) \left( {\mbox{\boldmath${                \theta}$}}-{%
\mbox{\boldmath${\bar{\theta}}$}}\right) ^{\prime }p\left( {\ %
\mbox{\boldmath${\vartheta}$}}|\mathbf{y}\right) d{%
\mbox{\boldmath${
\vartheta}$}}\right]^{-1}\right\}.
\end{eqnarray*}

Let $\{{\mbox{\boldmath${\vartheta}$}}^{\left( j\right) },j=1,2,\cdots ,J\}$
as the efficient random draws from $p\left( {\ \mbox{\boldmath${\vartheta}$}}%
|\mathbf{y}\right)$ and $\mathbf{A} = \left({\mbox{\boldmath${\bar\theta-%
\theta_{0}}$}}\right)\left({\mbox{\boldmath${\bar\theta-\theta_{0}}$}}%
\right)^\prime$. Then, we can get that
\begin{equation*}
\int \left( {\mbox{\boldmath${\theta}$}}-{\mbox{\boldmath${\bar{
\theta}}$}}\right) \left( {\mbox{\boldmath${\theta}$}}-{%
\mbox{\boldmath${
\bar{\theta}}$}}\right) ^{\prime }p\left( {\mbox{\boldmath${\vartheta}$} }|%
\mathbf{y}\right) \mbox{d}{\mbox{\boldmath${\vartheta}$}}\approx \widehat{{
\mbox{\boldmath${H}$}}}=\frac{1}{J}\sum_{j=1}^{J}\left( {\ %
\mbox{\boldmath${\theta}$}}^{\left( j\right) }-{%
\mbox{\boldmath${\bar{
\theta}}$}}\right) \left( {\mbox{\boldmath${\theta}$}}^{\left(j\right) }-{\ %
\mbox{\boldmath${\bar{\theta}}$}}\right) ^{\prime }=\frac{1}{J}\sum_{j=1}^{J}%
{\mbox{\boldmath${H}$}}^{\left(j\right) }.
\end{equation*}%
Then, we have
\begin{equation*}
\widehat{\mathbf{T}}\left( \mathbf{y},{\mbox{\boldmath${\theta}$}}
_{0}\right) = p + \mathbf{tr}\left(A\widehat{{\mbox{\boldmath${H}$}}}%
^{-1}\right).
\end{equation*}
which is the consistent estimator of $\mathbf{T}\left( \mathbf{y},{%
\mbox{\boldmath${\theta}$}} _{0}\right)$.

Following the notations of Magnus and Neudecker (2002) about matrix
derivatives, let
\begin{equation*}
{\mbox{\boldmath${h}$}}^{\left( j\right) }=vech\left( {\mbox{\boldmath${H}$}}%
^{\left(j\right) }\right) ,\ \widehat{\mathbf{h}}=vech\left(\widehat{{%
\mbox{\boldmath${H}$}}}\right).
\end{equation*}%
Note that the dimension of $\widehat{{\mbox{\boldmath${h}$}}}$ is $p^{\ast
}\times 1,p^{\ast }=p\left( p+1\right) /2$. Hence, we have
\begin{eqnarray*}
\frac{\partial \widehat{\mathbf{T}}\left( \mathbf{y},{%
\mbox{\boldmath${
\theta}$}}_{0}\right) }{\partial \widehat{{\mbox{\boldmath${h}$}}}} &=&
-vec(A^\prime)^{\prime}\left(\widehat{{\mbox{\boldmath${H}$}}}%
^{\prime-1}\otimes\widehat{{\mbox{\boldmath${H}$}}}^{-1}\right)\frac{
\partial \widehat{{\mbox{\boldmath${H}$}}}}{\partial \widehat{{\ %
\mbox{\boldmath${h}$}}}}
\end{eqnarray*}%
where
\begin{equation*}
\frac{ \partial \widehat{{\mbox{\boldmath${H}$}}}}{\partial \widehat{{\ %
\mbox{\boldmath${h}$}}}} = \left( \frac{\partial vec(\widehat{{%
\mbox{\boldmath${H}$}}})}{ \partial \widehat{{\mbox{\boldmath${h}$}}}}%
\right)_{p^{2}\times p^{\ast }} .
\end{equation*}

By the Delta method,
\begin{equation*}
Var\left( \widehat{\mathbf{T}}\left( \mathbf{y},{\mbox{\boldmath${\theta}$}}
_{0}\right) \right) =\frac{\partial \widehat{\mathbf{T}}\left( \mathbf{y},{\ %
\mbox{\boldmath${\theta}$}}_{0}\right) }{\partial \widehat{\mathbf{h}}}
Var\left( \widehat{{\mbox{\boldmath${h}$}}}\right) \left( \frac{\partial
\widehat{\mathbf{T}}\left( \mathbf{y},{\mbox{\boldmath${\theta}$}}
_{0}\right) }{\partial \widehat{{\mbox{\boldmath${h}$}}}}\right) ^{\prime }.
\end{equation*}%

\subsection{Proof of Theorem \ref{thm2}}
As in the the Proof of Theorem \ref{thm1}, under the null hypothesis,
\begin{equation*}
\widehat{\mbf{\vartheta}}-\widehat{\mbf{\vartheta}}_{m}=o_{p}\left(n^{-\frac{1}{2}}\right),\widehat{\mbf{\vartheta}}-\bar{\mbf{\vartheta}}=o_{p}\left(n^{-\frac{1}{2}}\right),\widehat{\mbf{\vartheta}}-\mbf{\vartheta}_{0}=O_{p}\left(n^{-\frac{1}{2}}\right)
\end{equation*}
and for the function $V\left(\tilde{\mbf{\vartheta}}\right)=E\left[\left(\mbf{\vartheta}-\tilde{\mbf{\vartheta}}\right)\left(\mbf{\vartheta}-\tilde{\mbf{\vartheta}}\right)^{\prime}|\mbf{y}\right]$,
\begin{equation*}
V\left(\bar{\mbf{\vartheta}}\right)=V\left(\widehat{\mbf{\vartheta}}_{m}\right)+o_{p}\left(n^{-1}\right)=-\ddot{\mathcal{L}}^{-1}\left(\widehat{\mbf{\vartheta}}\right)+o_{p}\left(n^{-1}\right),
\end{equation*}
\begin{equation*}
V\left(\mbf{\vartheta}_{0}\right)=V\left(\widehat{\mbf{\vartheta}}\right)+\left(\widehat{\mbf{\vartheta}}-\mbf{\vartheta}_{0}\right)\left(\widehat{\mbf{\vartheta}}-\mbf{\vartheta}_{0}\right)^{\prime}+o_{p}\left(n^{-1}\right),
\end{equation*}
\begin{equation*}
V\left(\bar{\mbf{\vartheta}}\right)=V\left(\widehat{\mbf{\vartheta}}\right)+o_{p}\left(n^{-1}\right).
\end{equation*}
We define
\begin{eqnarray*}
V_{R}\left(\tilde{\mbf{\vartheta}}\right) & =&E\left[\left(R\mbf{\vartheta}-R\tilde{\mbf{\vartheta}}\right)\left(R\mbf{\vartheta}-R\tilde{\mbf{\vartheta}}\right)^{\prime}|\mbf{y}\right]\\
 & =& RE\left[\left(\mbf{\vartheta}-\tilde{\mbf{\vartheta}}\right)\left(\mbf{\vartheta}-\tilde{\mbf{\vartheta}}\right)^{\prime}|\mbf{y}\right]R^{\prime}\\
 & =&RV\left(\tilde{\mbf{\vartheta}}\right)R^{\prime}.
\end{eqnarray*}
Then similarly, under the null hyposthesis, we have
\begin{equation*}
V_{R}\left(\bar{\mbf{\vartheta}}\right)=RV\left(\widehat{\mbf{\vartheta}}_{m}\right)R^{\prime}+o_{p}\left(n^{-1}\right)=-R\ddot{\mathcal{L}}^{-1}\left(\widehat{\mbf{\vartheta}}\right)R^{\prime}+o_{p}\left(n^{-1}\right),
\end{equation*}
\begin{equation*}
V_{R}\left(\mbf{\vartheta}_{0}\right)=RV\left(\widehat{\mbf{\vartheta}}\right)R^{\prime}+R\left(\widehat{\mbf{\vartheta}}-\mbf{\vartheta}_{0}\right)\left(\widehat{\mbf{\vartheta}}-\mbf{\vartheta}_{0}\right)^{\prime}R^{\prime}+o_{p}\left(n^{-1}\right),
\end{equation*}
\begin{equation*}
V_{R}\left(\bar{\mbf{\vartheta}}\right)=RV\left(\widehat{\mbf{\vartheta}}\right)R^{\prime}+o_{p}\left(n^{-1}\right).
\end{equation*}
Then the statistic can be rewritten as
\begin{eqnarray*}
\mbf{T}\left(\mbf{y},\mbf{r}\right) & =& \int\left(R\mbf{\vartheta}-\mbf{r}\right)^{\prime}\left[RV\left(\mbf{\bar{\vartheta}}\right)R^{\prime}\right]^{-1}\left(R\mathbf{\mbf{\vartheta}}-\mbf{r}\right)d\mbf{\vartheta}\\
 & =& tr\left\{ \left[\left[RV\left(\mbf{\bar{\vartheta}}\right)R^{\prime}\right]^{-1}\right]E\left[\left(R\mathbf{\mbf{\vartheta}}-\mbf{r}\right)\left(R\mathbf{\mbf{\vartheta}}-\mbf{r}\right)^{\prime}|\mbf{y}\right]\right\} \\
 & =&tr\left\{ \left(\left[RV\left(\widehat{\mbf{\vartheta}}\right)R^{\prime}\right]^{-1}+o_{p}\left(n\right)\right)V_{R}\left(\mbf{\vartheta}_{0}\right)\right\} \\
 & =&tr\left\{ \left[RV\left(\widehat{\mbf{\vartheta}}\right)R^{\prime}\right]^{-1}V_{R}\left(\mbf{\vartheta}_{0}\right)\right\} +tr\left\{ o_{p}\left(n\right)V_{R}\left(\mbf{\vartheta}_{0}\right)\right\} \\
 & =&tr\left\{ \left[RV\left(\widehat{\mbf{\vartheta}}\right)R^{\prime}\right]^{-1}\left[RV\left(\widehat{\mbf{\vartheta}}\right)R^{\prime}+R\left(\widehat{\mbf{\vartheta}}-\mbf{\vartheta}_{0}\right)\left(\widehat{\mbf{\vartheta}}-\mbf{\vartheta}_{0}\right)^{\prime}R^{\prime}+o_{p}\left(n^{-1}\right)\right]\right\} \\
 & &+o_{p}\left(n\right)O_{p}\left(n^{-1}\right)\\
 & =&m+tr\left\{ \left[RV\left(\widehat{\mbf{\vartheta}}\right)R^{\prime}\right]^{-1}R\left(\widehat{\mbf{\vartheta}}-\mbf{\vartheta}_{0}\right)\left(\widehat{\mbf{\vartheta}}-\mbf{\vartheta}_{0}\right)^{\prime}R^{\prime}\right\} +o_{p}\left(1\right)\\
 & =&m+\left(R\widehat{\mbf{\vartheta}}-r\right)^{\prime}\left[RV\left(\widehat{\mbf{\vartheta}}\right)R^{\prime}\right]^{-1}\left(R\widehat{\mbf{\vartheta}}-r\right)+o_{p}\left(1\right).
\end{eqnarray*}
Since it is obvious that $-\ddot{\mathcal{L}}^{-1}\left(\widehat{\mbf{\vartheta}}\right)=O_{p}\left(n^{-1}\right)$,
combined with $\widehat{\mbf{\vartheta}}-\mbf{\vartheta}_{0}=O_{p}\left(n^{-\frac{1}{2}}\right)$,
$V_{R}\left(\mbf{\vartheta}_{0}\right)=O_{p}\left(n^{-1}\right)+o_{p}\left(n^{-1}\right)+O_{p}\left(n^{-\frac{1}{2}}\right)O_{p}\left(n^{-\frac{1}{2}}\right)=O_{p}\left(n^{-1}\right)$.
And then
\begin{eqnarray*}
\left(R\widehat{\mbf{\vartheta}}-r\right)^{\prime}\left[RV\left(\widehat{\mbf{\vartheta}}\right)R^{\prime}\right]^{-1}\left(R\widehat{\mbf{\vartheta}}-r\right) & =&\left(R\widehat{\mbf{\vartheta}}-r\right)^{\prime}\left[-R\ddot{\mathcal{L}}^{-1}\left(\widehat{\mbf{\vartheta}}\right)R^{\prime}+o_{p}\left(n^{-1}\right)\right]^{-1}\left(R\widehat{\mbf{\vartheta}}-r\right)\\
 & =& \left(R\widehat{\mbf{\vartheta}}-r\right)^{\prime}\left[-R\ddot{\mathcal{L}}^{-1}\left(\widehat{\mbf{\vartheta}}\right)R^{\prime}\right]^{-1}\left(R\widehat{\mbf{\vartheta}}-r\right)\\
 && +o_{p}\left(n^{-1}\right)\left(R\widehat{\mbf{\vartheta}}-r\right)^{\prime}\left(R\widehat{\mbf{\vartheta}}-r\right)\\
 & =& \left(R\widehat{\mbf{\vartheta}}-r\right)^{\prime}\left[-R\ddot{\mathcal{L}}^{-1}\left(\widehat{\mbf{\vartheta}}\right)R^{\prime}\right]^{-1}\left(R\widehat{\mbf{\vartheta}}-r\right)+o_{p}\left(1\right)\\
 & =& \text{\textbf{Wald}}+o_{p}\left(1\right).
\end{eqnarray*}
Further, we know that under the null hypothesis, $R\mbf{\vartheta}_{0}=r$,
according to the standard maximum likelihood theory,
\begin{equation*}
\sqrt{n}\left(R\widehat{\mbf{\vartheta}}-r\right)=\sqrt{n}R\left(\widehat{\mbf{\vartheta}}-\mbf{\vartheta}_{0}\right)\overset{d}{\rightarrow}N\left(0,-nR\ddot{\mathcal{L}}^{-1}\left(\mbf{\vartheta}_{0}\right)R^{\prime}\right),
\end{equation*}
which implies that
\begin{equation*}
\text{\textbf{Wald}}=\left(R\widehat{\mbf{\vartheta}}-r\right)^{\prime}\left[-R\ddot{\mathcal{L}}^{-1}\left(\widehat{\mbf{\vartheta}}\right)R^{\prime}\right]^{-1}\left(R\widehat{\mbf{\vartheta}}-r\right)\overset{d}{\rightarrow}\chi^{2}\left(m\right).
\end{equation*}
Therefore,
\begin{equation*}
\mbf{T}\left(\mbf{y},\mbf{r}\right)=\text{\textbf{Wald}}+o_{p}\left(1\right)\overset{d}{\rightarrow}\chi^{2}\left(m\right).
\end{equation*}

\subsection{Proof of the Corollary \ref{corol2}}
Similar to the proof of Colorally \ref{corol1},
\begin{align*}
\mbf{T}\left(\mbf{y},\mbf{r}\right) & =\int\left(R\mbf{\vartheta}-\mbf{r}\right)^{\prime}\left[RV\left(\mbf{\bar{\vartheta}}\right)R^{\prime}\right]^{-1}\left(R\mathbf{\mbf{\vartheta}}-\mbf{r}\right)d\mbf{\vartheta}\\
 & =m+tr\left\{ \left(R\bar{\mbf{\vartheta}}-\mbf{r}\right)\left(R\bar{\mbf{\vartheta}}-\mbf{r}\right)^{\prime}\left[RV\left(\bar{\mbf{\vartheta}}\right)R^{\prime}\right]^{-1}\right\} \\
 & =m+tr\left\{ \mbf{A}\left[RV\left(\bar{\mbf{\vartheta}}\right)R^{\prime}\right]^{-1}\right\} .
\end{align*}
Let $\left\{ \mbf{\vartheta}^{\left(j\right)},j=1,\dots,J\right\} $
be the efficient random draws from $p\left(\mbf{\vartheta}|\mbf{y}\right)$,
then we can get that
\begin{equation*}
V\left(\bar{\mbf{\vartheta}}\right)=E\left[\left(\mbf{\vartheta}-\bar{\mbf{\vartheta}}\right)\left(\mbf{\vartheta}-\bar{\mbf{\vartheta}}\right)^{\prime}|\mbf{y}\right]\thickapprox\widehat{\mbf{H}}=\frac{1}{J}\sum_{j=1}^{J}\left(\mbf{\vartheta}^{\left(j\right)}-\bar{\mbf{\vartheta}}\right)\left(\mbf{\vartheta}^{\left(j\right)}-\bar{\mbf{\vartheta}}\right)^{\prime}.
\end{equation*}
The we have
\begin{equation*}
\widehat{\mbf{T}}\left(\mbf{y},\mbf{r}\right)=m+tr\left\{ \mbf{A}\left(R\widehat{\mbf{H}}R^{\prime}\right)^{-1}\right\} ,
\end{equation*}
therefore, let
\begin{equation*}
\mbf{h}^{\left(j\right)}=vech\left(\mbf{H}^{\left(j\right)}\right),\widehat{\mbf{h}}=vech\left(\widehat{\mbf{H}}\right),
\end{equation*}
then,
\begin{equation*}
\frac{\partial\widehat{\mbf{T}}\left(\mbf{y},\mbf{r}\right)}{\partial\widehat{\mbf{h}}}=-vec\left(\mbf{A}^{\prime}\right)^{\prime}\left[\left(R\widehat{\mbf{H}}R^{\prime}\right)^{-1}\otimes\left(R\widehat{\mbf{H}}R^{\prime}\right)^{-1}\right]\left(R\otimes R\right)\frac{\partial\widehat{\mbf{H}}}{\partial\widehat{\mbf{h}}},
\end{equation*}
where
\begin{equation*}
\frac{\partial\widehat{\mbf{H}}}{\partial\widehat{\mbf{h}}}=\left(\frac{\partial vec\left(\widehat{\mbf{H}}\right)}{\partial\widehat{\mbf{h}}}\right).
\end{equation*}
By the Delta method,
\begin{equation*}
Var\left(\widehat{\mbf{T}}\left(\mbf{y},\mbf{r}\right)\right)=\frac{\partial\widehat{\mbf{T}}\left(\mbf{y},\mbf{r}\right)}{\partial\widehat{\mbf{h}}}Var\left(\widehat{\mbf{h}}\right)\left(\frac{\partial\widehat{\mbf{T}}\left(\mbf{y},\mbf{r}\right)}{\partial\widehat{\mbf{h}}}\right)^{\prime}.
\end{equation*}

\subsection{Derivation of the statistics for	linear regression model} \label{smlexpl1}
Since the likelihood and the prior are both in Normal-Gamma form,
the intergretation of $\sigma^{2}$ gives the following result. \\
\begin{eqnarray*}
p\left(\mbf{\beta}|\mbf{y}\right) & \propto & \left[b+\frac{1}{2}\left(\mu_{0}^{\prime}V_{0}^{-1}\mu_{0}+\mbf{y}^{\prime}\mbf{y}-\mu^{*\prime}V^{*-1}\mu^{*}\right)+\frac{1}{2}\left(\mbf{\beta}-\mu^{*}\right)^{\prime}V^{*-1}\left(\mbf{\beta}-\mu^{*}\right)\right]^{-\frac{v+p}{2}}\\
 & \propto & \left[s+\frac{1}{2}\left(\mbf{\beta}-\mu^{*}\right)^{\prime}V^{*-1}\left(\mbf{\beta}-\mu^{*}\right)\right]^{-\frac{v+p}{2}}\\
 & \propto & \left[1+\frac{1}{v}\left(\mbf{\beta}-\mu^{*}\right)^{\prime}\left(\frac{2sV^{*}}{v}\right)^{-1}\left(\mbf{\beta}-\mu^{*}\right)\right]^{-\frac{v+p}{2}}
\end{eqnarray*}
where $v=2a+n$, $s=b+\frac{1}{2}\left(\mu_{0}^{\prime}V_{0}^{-1}\mu_{0}+\mbf{y}^{\prime}\mbf{y}-\mu^{*\prime}V^{*-1}\mu^{*}\right)$,
$V^{*}=\left(V_{0}^{-1}+\mbf{X}^{\prime}\mbf{X}\right)^{-1}$,$\mu^{*}=V^{*}\left(V_{0}^{-1}\tilde{\mu}+\mbf{X}^{\prime}\mbf{y}\right)$.
Then,
\begin{equation*}
\mbf{\beta}|\mbf{y}\sim t\left(\mu^{*},\frac{2sV^{*}}{v},v\right).
\end{equation*}
Hence, it is easy to get $\breve{\mbf{\beta}}|\mbf{y}\sim t\left(\breve{\mu}^{*},\frac{2s\breve{V}^{*}}{v},v\right)$,
where $\breve{\mu}^{*}$ is the subvector of $\mu^{*}$ corresponding
to $\breve{\mbf{\beta}}$ and $\breve{V}^{*}$ is similar.
Therefore, $Var\left(\mbf{\breve{\mbf{\beta}}}|\mbf{y}\right)=\frac{2s}{v-2}\breve{V}^{*}$.
Then the proposed statistics is
\begin{eqnarray*}
\mbf{T}\left(\mbf{y},\breve{\mbf{\beta}}_{0}\right) & = & \int\left(\breve{\mbf{\beta}}-\breve{\mbf{\beta}}_{0}\right)^{\prime}\left[Var\left(\mbf{\breve{\mbf{\beta}}}|\mbf{y}\right)\right]^{-1}\left(\breve{\mbf{\beta}}-\breve{\mbf{\beta}}_{0}\right)d\breve{\mbf{\beta}}\\
 & = & E\left\{ tr\left[\left(\breve{\mbf{\beta}}-\breve{\mbf{\beta}}_{0}\right)\left(\breve{\mbf{\beta}}-\breve{\mbf{\beta}}_{0}\right)^{\prime}Var\left(\mbf{\breve{\mbf{\beta}}}|\mbf{y}\right)^{-1}\right]\right\} \\
 & = & p+tr\left[\left(\bar{\breve{\mbf{\beta}}}-\breve{\mbf{\beta}}_{0}\right)\left(\bar{\breve{\mbf{\beta}}}-\breve{\mbf{\beta}}_{0}\right)^{\prime}Var\left(\mbf{\breve{\mbf{\beta}}}|\mbf{y}\right)^{-1}\right],
\end{eqnarray*}
where $\bar{\breve{\mbf{\beta}}}_{H_{1}}$ is the posterior
mean of $\breve{\mbf{\beta}}$ under $H_{1}$. Following the
result above, it can be simplified as
\begin{equation*}
\mbf{T}\left(\mbf{y},\breve{\mbf{\beta}}_{0}\right)=p+\frac{v-2}{2s}\left(\bar{\breve{\mbf{\beta}}}_{H_{1}}-\breve{\mbf{\beta}}_{0}\right)^{\prime}\breve{V}^{*-1}\left(\bar{\breve{\mbf{\beta}}}_{H_{1}}-\breve{\mbf{\beta}}_{0}\right),
\end{equation*}
where $p$ is the dimension of $\breve{\mbf{\beta}}$.

For the second hypothesis problem, the statistic can be derived readily, which is,
\begin{equation*}
\mbf{T}\left(\mbf{y},\mbf{r}\right)=m+\frac{v-2}{2s}\left(R\bar{\mbf{\beta}}_{H_{1}}-\mbf{r}\right)^{\prime}\left(RV^{*}R^{\prime}\right)^{-1}\left(R\bar{\mbf{\beta}}_{H_{1}}-\mbf{r}\right).
\end{equation*}

\subsection{Derivation of the statistics and Bayes factor for
Probit model} \label{empexpl1}

Let ${\mbox{\boldmath${\vartheta}$}}$ denotes all the parameters. And denote
${\mbox{\boldmath${\mu}$}}_{i}=\left(%
\begin{array}{c}
{\mbox{\boldmath${\beta}$}}_{h}^{\prime}{\mbox{\boldmath${x}$}}_{hi} \\
{\mbox{\boldmath${s}$}}_{i}^{\prime}{\mbox{\boldmath${\eta}$}}+{%
\mbox{\boldmath${\beta}$}}_{u}^{\prime}{\mbox{\boldmath${x}$}}_{ui}%
\end{array}%
\right)$,$\sigma_{h|u}^{2}=\left(1-\left(\frac{\sigma_{hu}}{%
\sigma_{h}\sigma_{u}}\right)^{2}\right)\sigma_{h}^{2}$, $\mu_{h|u}={%
\mbox{\boldmath${\beta}$}}_{h}^{\prime}{\mbox{\boldmath${x}$}}_{hi}+\frac{%
\sigma_{hu}}{\sigma_{u}^{2}}\left(y_{ui}-{\mbox{\boldmath${s}$}}_{i}^{\prime}%
{\mbox{\boldmath${\eta}$}}-{\mbox{\boldmath${\beta}$}}_{u}^{\prime}{%
\mbox{\boldmath${x}$}}_{ui}\right)$, then the log-likelihood is
\begin{eqnarray*}
\log p\left(Data|{\mbox{\boldmath${\vartheta}$}}\right) & = &
\sum_{i=1}^{N}\log\Phi\left(A_{i};{\mbox{\boldmath${\mu}$}}%
_{i},\Sigma\right)1_{\left\{ w_{ui}=0\right\}
}+\sum_{i=1}^{N}\log\Phi\left(B_{i};{\mbox{\boldmath${\mu}$}}%
_{i},\Sigma\right)1_{\left\{ \omega_{ui}=1\right\} } \\
& &
+\sum_{i=1}^{N}\log\Phi\left(C_{i};\mu_{h|u},\sigma_{h|u}^{2}\right)1_{\left%
\{ 0<\omega_{ui}<1\right\} } \\
& & +\sum_{i=1}^{N}\log\phi\left(y_{ui}|{\mbox{\boldmath${s}$}}_{i}^{\prime}{%
\mbox{\boldmath${\eta}$}}+{\mbox{\boldmath${\beta}$}}_{u}^{\prime}{%
\mbox{\boldmath${x}$}}_{ui},\sigma_{u}^{2}\right)1_{\left\{
0<\omega_{ui}<1\right\} },
\end{eqnarray*}
where $A_{i}=\left\{ \left(u,v\right):u\in\left[\gamma_{z_{hi}},%
\gamma_{z_{hi}+1}\right],v\in(-\infty,0]\right\} $, $B_{i}=\left\{
\left(u,v\right):u\in\left[\gamma_{z_{hi}},\gamma_{z_{hi}+1}\right]%
,v\in[1,+\infty)\right\} $, and $C_{i}=\left\{ u:u\in\left[%
\gamma_{z_{hi}},\gamma_{z_{hi}+1}\right]\right\} $.

Assume we want to test whether a subvector of ${\mbox{\boldmath${\beta}$}}%
_{h}^{\prime}$, ${\mbox{\boldmath${\theta}$}}={\mbox{\boldmath${\theta}$}}%
_{0}=0$ or not, that is,
\begin{equation*}
H_{0}:{\mbox{\boldmath${\theta}$}}={\mbox{\boldmath${\theta}$}}_{0},vs,H_{1}:%
{\mbox{\boldmath${\theta}$}}\neq{\mbox{\boldmath${\theta}$}}_{0}.
\end{equation*}
And the rest of the parameters is denoted by ${\mbox{\boldmath${\psi}$}}$, ${%
\mbox{\boldmath${\vartheta}$}}=\left({\mbox{\boldmath${\theta}$}}^{\prime},{%
\mbox{\boldmath${\psi}$}}^{\prime}\right)^{\prime}$.

\begin{itemize}
\item The estimator of ${\mbox{\boldmath${T}$}}\left(Data,{%
\mbox{\boldmath${\theta}$}}_{0}\right)$  and its NSE.\newline
Let $\left\{ {\mbox{\boldmath${\theta}$}}^{\left(j\right)}\right\} _{j=1}^{J}
$  denote the effective posterior draws of the targeted parameter $\tilde{%
\beta}$.  The statistic can be calculated as
\begin{equation*}
\widehat{{\mbox{\boldmath${T}$}}}\left(Data,{\mbox{\boldmath${\theta}$}}%
_{0}\right)=\frac{1}{J}\sum_{j=1}^{J}\left({\mbox{\boldmath${\theta}$}}%
^{\left(j\right)}-{\mbox{\boldmath${\theta}$}}_{0}\right)^{\prime}\widehat{H}%
^{-1}\left({\mbox{\boldmath${\theta}$}}^{\left(j\right)}-{%
\mbox{\boldmath${\theta}$}}_{0}\right).
\end{equation*}
where $\widehat{{\mbox{\boldmath${H}$}}}=\frac{1}{J}\sum_{j=1}^{J}\left({%
\mbox{\boldmath${\theta}$}}^{\left(j\right)}-\bar{{\mbox{\boldmath${\theta}$}%
}}\right)\left({\mbox{\boldmath${\theta}$}}^{\left(j\right)}-\bar{{%
\mbox{\boldmath${\theta}$}}}\right)^{\prime}=\frac{1}{J}\sum_{j=1}^{J}{%
\mbox{\boldmath${H}$}}^{\left(j\right)}$,  $\bar{{\mbox{\boldmath${\theta}$}}%
}=\frac{1}{J}\sum_{j=1}^{J}{\mbox{\boldmath${\theta}$}}^{\left(j\right)}$.
And the corresponding numerical standard error is
\begin{equation*}
{\mbox{\boldmath${h}$}}^{\left(j\right)}=vech\left({\mbox{\boldmath${H}$}}%
^{\left(j\right)}\right),\widehat{{\mbox{\boldmath${h}$}}}=vech\left(%
\widehat{{\mbox{\boldmath${H}$}}}\right),
\end{equation*}
\begin{equation*}
{\mbox{\boldmath${A}$}}=\frac{1}{J}\sum_{j=1}^{J}\left({\mbox{\boldmath${%
\theta}$}}^{\left(j\right)}-{\mbox{\boldmath${\theta}$}}_{0}\right)\left({%
\mbox{\boldmath${\theta}$}}^{\left(j\right)}-{\mbox{\boldmath${\theta}$}}%
_{0}\right)^{\prime},
\end{equation*}

\begin{equation*}
\frac{\partial\widehat{{\mbox{\boldmath${T}$}}}\left(Data,{%
\mbox{\boldmath${\theta}$}}_{0}\right)}{\partial\widehat{{%
\mbox{\boldmath${h}$}}}^{\prime}}=-vec\left({\mbox{\boldmath${A}$}}%
^{\prime}\right)^{\prime}\left(\widehat{{\mbox{\boldmath${H}$}}}^{-1}\otimes%
\widehat{{\mbox{\boldmath${H}$}}}^{-1}\right)\frac{\partial vec\left(%
\widehat{{\mbox{\boldmath${H}$}}}\right)}{\partial\widehat{{%
\mbox{\boldmath${h}$}}}^{\prime}},
\end{equation*}
\begin{equation*}
Var\left({\widehat{{\mbox{\boldmath${T}$}}}\left(Data,{\mbox{\boldmath${%
\theta}$}}_{0}\right)}\right)=\frac{\partial\widehat{{\mbox{\boldmath${T}$}}}%
\left(Data,{\mbox{\boldmath${\theta}$}}_{0}\right)}{\partial\widehat{{%
\mbox{\boldmath${h}$}}}^{\prime}}Var\left(\widehat{{\mbox{\boldmath${h}$}}}%
\right)\frac{\partial\widehat{{\mbox{\boldmath${T}$}}}\left(Data,{%
\mbox{\boldmath${\theta}$}}_{0}\right)}{\partial\widehat{{%
\mbox{\boldmath${h}$}}}},
\end{equation*}
where
\begin{equation*}
Var\left(\widehat{{\mbox{\boldmath${h}$}}}\right)=\frac{1}{J}\left[%
\Omega_{0}+\sum_{k=1}^{q}\left(1-\frac{k}{q+1}\right)\left(\Omega_{k}+%
\Omega_{k}^{\prime}\right)\right],
\end{equation*}
with
\begin{equation*}
\Omega_{k}=\frac{1}{J}\sum_{j=k+1}^{J}\left({\mbox{\boldmath${h}$}}%
^{\left(j\right)}-\widehat{{\mbox{\boldmath${h}$}}}\right)\left({%
\mbox{\boldmath${h}$}}^{\left(j\right)}-\widehat{{\mbox{\boldmath${h}$}}}%
\right)^{\prime},
\end{equation*}
and the value of $q$ is 10.

\item The estimator of LLY statistic and its NSE.\newline
When the last two terms are equal to 1 for each $i$, let ${%
\mbox{\boldmath${y}$}}_{i}=\left(y_{hi},y_{ui}\right)^{\prime}$,  ${%
\mbox{\boldmath${x}$}}_{i}=\left(%
\begin{array}{cc}
x_{hi}^{\prime} & 0 \\
0 & x_{ui}^{\prime}%
\end{array}%
\right)$, by the Leibnitz's rule, the first derivative of the log-likelihood
with respect to ${\mbox{\boldmath${\theta}$}}$ is
\begin{eqnarray*}
& & \frac{\partial\log p\left(Data|{\mbox{\boldmath${\vartheta}$}}\right)}{%
\partial{\mbox{\boldmath${\theta}$}}} \\
& = & \sum_{i=1}^{N}\frac{{\mbox{\boldmath${x}$}}_{i}^{{\mbox{\boldmath${%
\theta}$}}}}{\Phi\left(A_{i};{\mbox{\boldmath${\mu}$}}_{i},\Sigma\right)}%
\int_{-\infty}^{-\frac{\mu_{u}}{\sigma_{u}}}\left[\phi\left(\frac{%
\gamma_{y_{hi}+1}-\mu_{h|u}}{\sigma_{h|u}}\right)-\phi\left(\frac{%
\gamma_{y_{hi}}-\mu_{h|u}}{\sigma_{h|u}}\right)\right]\phi\left(\frac{%
y_{ui}-\mu_{u}}{\sigma_{u}}\right)dy_{ui}1_{\left\{ w_{ui}=0\right\} } \\
& & +\sum_{i=1}^{N}\frac{x_{i}^{{\mbox{\boldmath${\theta}$}}}}{%
\Phi\left(B_{i};{\mbox{\boldmath${\mu}$}}_{i},\Sigma\right)}\int_{1-\frac{%
\mu_{u}}{\sigma_{u}}}^{+\infty}\left[\phi\left(\frac{\gamma_{y_{hi}+1}-%
\mu_{h|u}}{\sigma_{h|u}}\right)-\phi\left(\frac{\gamma_{y_{hi}}-\mu_{h|u}}{%
\sigma_{h|u}}\right)\right]\phi\left(\frac{y_{ui}-\mu_{u}}{\sigma_{u}}%
\right)dy_{ui}1_{\left\{ \omega_{ui}=1\right\} } \\
& & +\sum_{i=1}^{N}\frac{x_{i}^{{\mbox{\boldmath${\theta}$}}}}{%
\Phi\left(C_{i};\mu_{h|u},\sigma_{h|u}^{2}\right)}\left[\phi\left(\frac{%
\gamma_{y_{hi}+1}-\mu_{h|u}}{\sigma_{h|u}}\right)-\phi\left(\frac{%
\gamma_{y_{hi}}-\mu_{h|u}}{\sigma_{h|u}}\right)\right]1_{\left\{
0<\omega_{ui}<1\right\} }
\end{eqnarray*}
where $x_{i}^{{\mbox{\boldmath${\theta}$}}}$ is the explanary variables in  $%
x_{i}$ corresponding to $\tilde{\beta}$. Then
\begin{equation*}
C_{{\mbox{\boldmath${\theta}$}}{\mbox{\boldmath${\theta}$}}}\left(\bar{{%
\mbox{\boldmath${\vartheta}$}}_{0}}\right)=\left.\left(\frac{\partial\log
p\left(Data|{\mbox{\boldmath${\vartheta}$}}\right)}{\partial{%
\mbox{\boldmath${\theta}$}}}\right)\left(\frac{\partial\log p\left(Data|{%
\mbox{\boldmath${\vartheta}$}}\right)}{\partial{\mbox{\boldmath${\theta}$}}}%
\right)^{\prime}\right|_{{\mbox{\boldmath${\vartheta}$}}=\bar{{%
\mbox{\boldmath${\vartheta}$}}}_{0}},
\end{equation*}
where $\bar{{\mbox{\boldmath${\vartheta}$}}}_{0}=\left({\mbox{\boldmath${%
\theta}$}}_{0}^{\prime},\bar{{\mbox{\boldmath${\psi}$}}}_{0}^{\prime}%
\right)^{\prime}$  and $\bar{{\mbox{\boldmath${\psi}$}}}_{0}$ is the
posterior mean of ${\mbox{\boldmath${\psi}$}}$  under $H_{0}$. \newline
We firstly draw MCMC samples for the model under $H_{0}$ and calculate  $C_{{%
\mbox{\boldmath${\theta}$}}{\mbox{\boldmath${\theta}$}}}\left(\bar{{%
\mbox{\boldmath${\vartheta}$}}_{0}}\right)$.  After that, we run the MCMC
and obtain the samples of ${\mbox{\boldmath${\vartheta}$}}$  under $H_{1}$,
denoted as $\left\{ {\mbox{\boldmath${\vartheta}$}}^{\left(j\right)}\right\}
_{j=1}^{J}=\left\{ {\mbox{\boldmath${\theta}$}}^{\left(j\right)},{%
\mbox{\boldmath${\psi}$}}^{\left(j\right)}\right\} _{j=1}^{J}.$  Then LLY
statistic can be calculated by
\begin{equation*}
\widehat{{\mbox{\boldmath${T}$}}}_{LLY}\left(Data,{\mbox{\boldmath${\theta}$}%
}_{0}\right)=\frac{1}{J}\sum_{j=1}^{J}g\left({\mbox{\boldmath${\theta}$}}%
^{\left(j\right)}\right),
\end{equation*}
where
\begin{equation*}
g\left({\mbox{\boldmath${\theta}$}}^{\left(j\right)}\right)=\left({%
\mbox{\boldmath${\theta}$}}^{\left(j\right)}-\bar{{\mbox{\boldmath${\theta}$}%
}}\right)^{\prime}C\left(\bar{{\mbox{\boldmath${\vartheta}$}}_{0}}%
\right)\left({\mbox{\boldmath${\theta}$}}^{\left(j\right)}-\bar{{%
\mbox{\boldmath${\theta}$}}}\right).
\end{equation*}
Then the numerical variance of $\widehat{{\mbox{\boldmath${T}$}}}%
_{LLY}\left(Data,{\mbox{\boldmath${\theta}$}}_{0}\right)$  is
\begin{equation*}
Var\left(\widehat{{\mbox{\boldmath${T}$}}}_{LLY}\left(Data,{%
\mbox{\boldmath${\theta}$}}_{0}\right)\right)=\frac{1}{J}\left\{
\Omega_{0}+\sum_{k=1}^{q}\left(1-\frac{k}{q+1}\right)\left(\Omega_{k}+%
\Omega_{k}^{\prime}\right)\right\} ,
\end{equation*}
where
\begin{equation*}
\Omega_{k}=\frac{1}{J}\sum_{j=k+1}^{J}\left(g\left({\mbox{\boldmath${%
\theta}$}}^{\left(j\right)}\right)-\widehat{{\mbox{\boldmath${T}$}}}%
_{LLY}\left(Data,{\mbox{\boldmath${\theta}$}}_{0}\right)\right)^{2}.
\end{equation*}

\item The Bayes factor and the corresponding NSE.\newline
Following Chib(1995), the logarithmic marginal likelihood under $H_{1}$,$%
\log p(y|M1)$,  is given by
\begin{equation*}
\log p\left(Data|H_{1}\right)=\log p\left(Data|\bar{{\mbox{\boldmath${%
\vartheta}$}}}\right)+\log p\left(\bar{{\mbox{\boldmath${\vartheta}$}}}%
\right)-\log p\left(\bar{{\mbox{\boldmath${\vartheta}$}}}|Data\right),
\end{equation*}
where $\log p\left(Data|\bar{{\mbox{\boldmath${\vartheta}$}}}\right)$ is
known.  $p\left(\bar{{\mbox{\boldmath${\vartheta}$}}}\right)$ is the p.d.f.
of the  prior evaluated at $\bar{{\mbox{\boldmath${\vartheta}$}}}$, $p\left(%
\bar{{\mbox{\boldmath${\vartheta}$}}}|y\right)$  is the p.d.f. of the
posterior distribution evaluated at $\bar{{\mbox{\boldmath${\vartheta}$}}}$.
The posterior quantity can be calculated by
\begin{equation*}
\widehat{p}\left(\bar{{\mbox{\boldmath${\vartheta}$}}}|Data\right)=\frac{1}{J%
}\sum_{j=1}^{G}p\left(\bar{{\mbox{\boldmath${\vartheta}$}}}|{%
\mbox{\boldmath${L}$}}_{1}^{\left(j\right)}\right),
\end{equation*}
where $\left\{ {\mbox{\boldmath${L}$}}_{1}^{\left(j\right)}\right\}
_{j=1}^{J}$  are the efficient draws of the latent variables from $p\left({%
\mbox{\boldmath${L}$}}_{1}|Data,\bar{{\mbox{\boldmath${\vartheta}$}}}\right)$%
.  For this specific model, $p\left({\mbox{\boldmath${\vartheta}$}}%
|Data\right)$  has analytical form. Therefore we can obtain the
approximation of  $\log p\left(Data|H_{1}\right)$, $\log\widehat{ p}%
\left(Data|H_{1}\right)$.  Similary, we can also approximate the logarithmic
marginal likelihood  under $H_{0}$, $\log p\left(Data|H_{0}\right)$.
\begin{equation*}
\log p\left(Data|H_{0}\right)=\log p\left(Data|\bar{{\mbox{\boldmath${%
\vartheta}$}}}_{0}\right)+\log p\left(\bar{{\mbox{\boldmath${\psi}$}}}%
_{0}\right)-\log p\left(\bar{{\mbox{\boldmath${\psi}$}}}_{0}|Data,{%
\mbox{\boldmath${\theta}$}}_{0}\right).
\end{equation*}
Similarly, $\widehat{p}\left(\bar{{\mbox{\boldmath${\psi}$}}}_{0}|Data,{%
\mbox{\boldmath${\theta}$}}_{0}\right)=\frac{1}{J}\sum_{j=1}^{J}p\left(\bar{{%
\mbox{\boldmath${\psi}$}}}_{0}|{\mbox{\boldmath${L}$}}_{0}^{\left(j\right)},{%
\mbox{\boldmath${\theta}$}}_{0}\right)$,  and $\left\{ {\mbox{\boldmath${L}$}%
}_{0}^{\left(j\right)}\right\} _{j=1}^{J}$  denotes the efficient draws from
$p\left({\mbox{\boldmath${L}$}}_{0}|Data,\bar{{\mbox{\boldmath${\vartheta}$}}%
}_{0}\right)$.  Therefore, the logarithmic Bayes factor can be estimated by
\begin{eqnarray*}
\widehat{\log BF}_{10} & = & \left[\log p\left(Data|\bar{{%
\mbox{\boldmath${\vartheta}$}}}\right)+\log p\left(\bar{{\mbox{\boldmath${%
\vartheta}$}}}\right)-\log\widehat{p}\left(\bar{{\mbox{\boldmath${%
\vartheta}$}}}|Data\right)\right] \\
& & -\left[\log p\left(Data|\bar{{\mbox{\boldmath${\vartheta}$}}}%
_{0}\right)+\log p\left(\bar{{\mbox{\boldmath${\psi}$}}}_{0}\right)-\log%
\widehat{p}\left(\bar{{\mbox{\boldmath${\psi}$}}}_{0}|Data\right)\right].
\end{eqnarray*}
To calculate the NSE, following Chib(1995), let $h_{1}^{\left(j\right)}=p%
\left(\bar{{\mbox{\boldmath${\vartheta}$}}}|{\mbox{\boldmath${L}$}}%
_{1}^{\left(j\right)}\right)$,  $h_{0}^{\left(j\right)}=p\left(\bar{{%
\mbox{\boldmath${\psi}$}}}_{0}|{\mbox{\boldmath${L}$}}_{0}^{\left(j\right)},{%
\mbox{\boldmath${\theta}$}}_{0}\right)$,  ${\mbox{\boldmath${h}$}}%
^{\left(j\right)}=\left(h_{1}^{\left(j\right)},h_{0}^{\left(j\right)}%
\right)^{\prime}$,  $\widehat{{\mbox{\boldmath${h}$}}}=\left(\widehat{h}_{1},%
\widehat{h}_{0}\right)^{\prime}$,  $\widehat{h}_{0}=\frac{1}{J}%
\sum_{j=1}^{J}h_{0}^{\left(j\right)}$, $\widehat{h}_{1}=\frac{1}{J}%
\sum_{j=1}^{J}h_{1}^{\left(j\right)}$.  Then the numerical variance is
\begin{equation*}
Var\left(\widehat{\log BF}_{10}\right)=\left(\frac{\partial\widehat{\log BF}%
_{10}}{\partial\widehat{{\mbox{\boldmath${h}$}}}}\right)^{\prime}Var\left(%
\widehat{{\mbox{\boldmath${h}$}}}\right)\left(\frac{\partial\widehat{\log BF}%
_{10}}{\partial\widehat{{\mbox{\boldmath${h}$}}}}\right),
\end{equation*}
\begin{equation*}
Var\left(\widehat{{\mbox{\boldmath${h}$}}}\right)=\frac{1}{J}\left[%
\Omega_{0}+\sum_{k=1}^{q}\left(1-\frac{k}{q+1}\right)\left(\Omega_{k}+%
\Omega_{k}^{\prime}\right)\right],
\end{equation*}
\begin{equation*}
\Omega_{k}=\frac{1}{J}\sum_{j=k+1}^{J}\left({\mbox{\boldmath${h}$}}%
^{\left(j\right)}-\widehat{{\mbox{\boldmath${h}$}}}\right)\left({%
\mbox{\boldmath${h}$}}^{\left(j\right)}-\widehat{{\mbox{\boldmath${h}$}}}%
\right)^{\prime},
\end{equation*}
\begin{equation*}
\frac{\partial\widehat{\log BF}_{10}}{\partial\widehat{{\mbox{\boldmath${h}$}%
}}}=\left(%
\begin{array}{c}
-\widehat{p}\left(\bar{{\mbox{\boldmath${\vartheta}$}}}|Data\right)^{-1} \\
\widehat{p}\left(\bar{{\mbox{\boldmath${\psi}$}}}_{0}|Data,{%
\mbox{\boldmath${\theta}$}}_{0}\right)^{-1}%
\end{array}%
\right),
\end{equation*}
where $q$ is always chosen as 10 in the literature.
\end{itemize}

\subsection{Derivation of the statistics and Bayes factor for
leverage stochastic volatility model}
\label{empexpl1}
\begin{itemize}
\item The estimator of ${\mbox{\boldmath${T}$}}\left({\mbox{\boldmath${y}$}}%
, {\mbox{\boldmath${\theta}$}}_{0}\right)$ and its NSE\newline
The proposed statistic is
\begin{equation*}
{\mbox{\boldmath${T\left(y,\theta_{0}\right)}$}}=\frac{\int\rho^{2}p\left(
\rho|{\mbox{\boldmath${y}$}}\right)d\rho}{\int\left(\rho-\bar{\rho}
\right)^{2}p\left(\rho|{\mbox{\boldmath${y}$}}\right)d\rho}\approx\frac{
\frac{1}{J}\sum_{j=1}^{J}\left(\rho^{\left(j\right)}\right)^{2}}{\frac{1}{J}
\sum_{j=1}^{J}\left(\rho^{\left(j\right)}-\bar{\rho}\right)^{2}}=\frac{
\widehat{d}_{1}}{\widehat{d}_{2}},
\end{equation*}
where $\rho^{\left(j\right)}$ is the $j$th effective draws of $\rho$ under $
H_{1}$, $\bar{\rho}$ is the posterior mean of $\rho$ under $H_{1}$, $
\widehat{d}_{1}=\frac{1}{J}\sum_{j=1}^{J}\left(\rho^{\left(j\right)}
\right)^{2}$ and $\widehat{d}_{2}=\frac{1}{J}\sum_{j=1}^{J}d_{2}^{\left(j
\right)}=\frac{1}{J}\sum_{j=1}^{J}\left(\rho^{\left(j\right)}-\bar{\rho}
\right)^{2}$.\newline
The NSE of the estimator can be obtained by
\begin{equation*}
NSE\left({\mbox{\boldmath${\widehat{T}\left(y,\theta_{0}\right)}$}}\right)=
\sqrt{\frac{\bar{\rho}^{2}}{\widehat{d}_{2}^{2}}Var\left(\widehat{d}
_{2}\right)},
\end{equation*}
\begin{equation*}
Var\left(\rho\right)=\frac{1}{J}\left[\Omega_{0}+\sum_{k=1}^{q}\left(1-\frac{
k}{q+1}\right)\left(\Omega_{k}+\Omega_{k}^{\prime}\right)\right]
\,\,\Omega_{k}=\frac{1}{J}\sum_{j=k+1}^{J}\left(d_{2}^{\left(j\right)}-
\widehat{d}_{2}\right)^{2},k=1,\dots,q.
\end{equation*}

\item The estimator of LLY statistic and its NSE\newline
For the likelihood, by introducing $\omega_{t}\sim N\left(0,1\right)$ and $
\omega_{t}$ is independent of $\varepsilon_{t+1}$, then $\epsilon_{t}=\sqrt{
1-\rho^{2}}\omega_{t}+\rho\varepsilon_{t+1}$, we rewite the model as
\begin{equation*}
\begin{cases}
y_{t}|h_{t},h_{t+1}=\frac{\rho}{\sigma}\exp\left(\frac{1}{2}h_{t}\right) %
\left[h_{t+1}-\mu-\phi\left(h_{t}-\mu\right)\right]+\exp\left(\frac{1}{2}
h_{t}\right)\sqrt{1-\rho^{2}}\omega_{t} & \omega_{t}\sim N\left(0,1\right)
\\
h_{t+1}|h_{t},\mu,\sigma,\phi=\mu+\phi\left(h_{t}-\mu\right)+\sigma
\varepsilon_{t+1} & \epsilon\varepsilon_{t+1}\sim N\left(0,1\right)%
\end{cases}
,
\end{equation*}
where $\epsilon_{t}$ and $\omega_{t}$ are independent. Hence, let ${\ %
\mbox{\boldmath${\vartheta}$}}=\left(\mu,\phi,\sigma^{-2},\rho\right)^{
\prime}$, the log-likelihood of data ${\mbox{\boldmath${y}$}}=\left\{
y_{t}\right\} _{t=1}^{n}$, $\log p\left({\mbox{\boldmath${y}$}}|{\ %
\mbox{\boldmath${h}$}},{\mbox{\boldmath${\vartheta}$}}\right)$, is
\begin{eqnarray*}
\log p\left({\mbox{\boldmath${y}$}}|{\mbox{\boldmath${h}$}},{\ %
\mbox{\boldmath${\vartheta}$}}\right) & = & \sum_{t=1}^{n}\left[-\frac{1}{
2\exp\left(h_{t}\right)\left(1-\rho^{2}\right)}\left(y_{t}-\frac{\rho}{%
\sigma }\exp\left(\frac{1}{2}h_{t}\right)\left(h_{t+1}-\mu-\phi\left(h_{t}-%
\mu \right)\right)\right)^{2}\right] \\
& & +\frac{n}{2}\log\left(2\pi\right)-\frac{1}{2}\sum_{t=1}^{n}h_{t}-\frac{n
}{2}\log\left(1-\rho^{2}\right).
\end{eqnarray*}
Thus the first derivative with respect to $\rho$ is
\begin{eqnarray*}
\frac{\partial\log p\left({\mbox{\boldmath${y}$}}|{\mbox{\boldmath${h}$}},{\ %
\mbox{\boldmath${\vartheta}$}}\right)}{\partial\rho} & = & -\frac{\rho}{
\left(1-\rho^{2}\right)^{2}}\sum_{t=1}^{n}\frac{y_{t}^{2}}{
\exp\left(h_{t}\right)} \\
& & +\frac{1+\rho^{2}}{\left(1-\rho^{2}\right)^{2}}\sum_{t=1}^{n}\frac{y_{t}
}{\sigma}\exp\left(-\frac{1}{2}h_{t}\right)\left(h_{t+1}-\mu-\phi
\left(h_{t}-\mu\right)\right) \\
& & -\frac{\rho}{\left(1-\rho^{2}\right)^{2}}\sum_{t=1}^{n}\frac{1}{
\sigma^{2}}\left(h_{t+1}-\mu-\phi\left(h_{t}-\mu\right)\right)^{2}+\frac{
n\rho}{1-\rho^{2}} \\
& = & \frac{\rho}{\left(1-\rho^{2}\right)^{2}}A+\frac{1+\rho^{2}}{
\left(1-\rho^{2}\right)^{2}}B+\frac{\rho n}{1-\rho^{2}}
\end{eqnarray*}
where $A=-\sum_{t=1}^{n}\frac{y_{t}^{2}}{\exp\left(h_{t}\right)}
-\sum_{t=1}^{n}\frac{1}{\sigma^{2}}\left(h_{t+1}-\mu-\phi\left(h_{t}-\mu
\right)\right)^{2}$, and similarly, $B=\sum_{t=1}^{n}\frac{1}{\sigma}
\exp\left(-\frac{1}{2}h_{t}\right)\left(h_{t+1}-\mu-\phi\left(h_{t}-\mu
\right)\right)y_{t}$. In order to calculate the statistic of LLY, the
observed first derivative function evaluated at the posterior mean ${\ %
\mbox{\boldmath${\bar{\vartheta}}$}}_{0}=\left(\bar{\mu}_{0},\bar{\phi}_{0},
\bar{\sigma}_{0}^{-2},0\right)^{\prime}$, under $H_{0}$, is,
\begin{eqnarray*}
s_{{\mbox{\boldmath${\theta}$}}}\left(\bar{{\mbox{\boldmath${\vartheta}$}}}
_{0}\right) & = & \left.\frac{\partial\log p\left({\mbox{\boldmath${y}$}},{\ %
\mbox{\boldmath${\vartheta}$}}\right)}{\partial\rho}\right|_{{\ %
\mbox{\boldmath${\vartheta=\bar{\vartheta}}$}}_{0}} \\
& = & \left.\frac{\partial\log p\left({\mbox{\boldmath${y}$}}|{\ %
\mbox{\boldmath${\vartheta}$}}\right)}{\partial\rho}\right|_{{\ %
\mbox{\boldmath${\vartheta=\bar{\vartheta}}$}}_{0}}+\left.\frac{\partial\log
p\left({\mbox{\boldmath${\vartheta}$}}\right)}{\partial\rho}\right|_{{\ %
\mbox{\boldmath${\vartheta=\bar{\vartheta}}$}}_{0}} \\
& = & \int\left.\frac{\partial\log p\left({\mbox{\boldmath${y}$}}|{\ %
\mbox{\boldmath${h}$}},{\mbox{\boldmath${\vartheta}$}}\right)}{\partial\rho}
p\left({\mbox{\boldmath${h}$}}|{\mbox{\boldmath${y}$}},{\mbox{%
\boldmath${					\vartheta}$}}\right)\right|_{{\mbox{\boldmath${\vartheta=%
\bar{\vartheta}}$}} _{0}}d{\mbox{\boldmath${h}$}}+\frac{\partial\log p\left({%
\mbox{\boldmath${						\vartheta}$}}\right)}{\partial\rho} \\
& \approx & \frac{1}{J}\sum_{j=1}^{J}\left.\frac{\partial\log p\left({\ %
\mbox{\boldmath${y}$}}|{\mbox{\boldmath${h}$}}^{\left(j\right)},{\ %
\mbox{\boldmath${\vartheta}$}}\right)}{\partial\rho}\right|_{{\ %
\mbox{\boldmath${\vartheta=\bar{\vartheta}}$}}_{0}}+\frac{\partial\log
p\left(\rho\right)}{\partial\rho} \\
& = & \frac{1}{J}\sum_{j=1}^{J}\left.\left[\frac{\rho}{\left(1-\rho^{2}
\right)^{2}}A^{\left(j\right)}+\frac{1+\rho^{2}}{\left(1-\rho^{2}\right)^{2}}
B^{\left(j\right)}+\frac{\rho n}{1-\rho^{2}}\right]\right|_{{\ %
\mbox{\boldmath${\vartheta}$}}=\bar{{\mbox{\boldmath${\vartheta}$}}}_{0}} \\
& = & \frac{1}{J}\sum_{j=1}^{J}B^{\left(j\right)} \\
& = & \widehat{d}_{3}.
\end{eqnarray*}
where $A_{t}^{\left(j\right)}=-\sum_{t=1}^{n}\frac{y_{t}^{2}}{
\exp\left(h_{t}^{\left(j\right)}\right)}-\sum_{t=1}^{n}\frac{1}{\sigma^{2}}
\left(h_{t+1}^{\left(j\right)}-\mu-\phi\left(h_{t}^{\left(j\right)}-\mu
\right)\right)^{2}$, and $B_{t}^{\left(j\right)}=\sum_{t=1}^{n}\frac{1}{\bar{
\sigma}_{0}}\exp\left(-\frac{1}{2}h_{t}^{\left(j\right)}\right)
\left(h_{t+1}^{\left(j\right)}-\bar{\mu}_{0}-\bar{\phi}_{0}\left(h_{t}^{
\left(j\right)}-\bar{\mu}_{0}\right)\right)y_{t}$, ${\mbox{\boldmath${h}$}}
^{\left(g\right)}=\left\{ h_{t}^{\left(g\right)}\right\} _{t=1}^{n}$ the $g$
th MCMC outputs of ghe latent variable ${\mbox{\boldmath${h}$}}$ from $
p\left({\mbox{\boldmath${h}$}}|{\mbox{\boldmath${y}$}},\bar{{\ %
\mbox{\boldmath${\vartheta}$}}}_{0}\right)$.\newline
Then the statistic of LLY is approximated by,
\begin{equation*}
\widehat{{\mbox{\boldmath${T}$}}}\left({\mbox{\boldmath${y}$}},{\ %
\mbox{\boldmath${\theta}$}}_{0}\right)\approx\widehat{d}_{3}^{2}\frac{1}{J}
\sum_{j=1}^{J}\left(\rho^{\left(j\right)}-\bar{\rho}\right)^{2}=\widehat{d}
_{3}^{2}\frac{1}{G}\sum_{j=1}^{J}d_{2}^{\left(j\right)}=\widehat{d}_{1}^{2}
\widehat{d}_{2},
\end{equation*}
where $d_{2}^{\left(j\right)}=\left(\rho^{\left(j\right)}-\bar{\rho}
\right)^{2}$ and $\rho^{\left(j\right)}$ is the $g$th MCMC output of
parameter $\rho$ under $H_{1}$. The first derivative of $\widehat{{\ %
\mbox{\boldmath${T}$}}}\left({\mbox{\boldmath${y}$}},{\mbox{\boldmath${				%
\theta}$}}_{0}\right)$ with respect to $\widehat{{\mbox{\boldmath${d}$}}}
=\left(\widehat{d}_{3},\widehat{d}_{2}\right)$ is
\begin{equation*}
\frac{\partial\widehat{{\mbox{\boldmath${T}$}}}\left({\mbox{\boldmath${y}$}}%
, {\mbox{\boldmath${\theta}$}}_{0}\right)}{\partial{\mbox{\boldmath${d}$}}}
=\left(
\begin{array}{cc}
2\widehat{d}_{3}\widehat{d}_{2}, & \widehat{d}_{3}^{2}%
\end{array}
\right).
\end{equation*}
And the corresponding standard error estimator is
\begin{equation*}
NSE\left(\widehat{{\mbox{\boldmath${T}$}}}\left({\mbox{\boldmath${y}$}},{\ %
\mbox{\boldmath${\theta}$}}_{0}\right)\right)=\sqrt{Var\left(\widehat{{\ %
\mbox{\boldmath${T}$}}}\left({\mbox{\boldmath${y}$}},{\mbox{%
\boldmath${						\theta}$}}_{0}\right)\right)}=\sqrt{\frac{\partial\widehat{{%
\ \mbox{\boldmath${T}$}}}\left({\mbox{\boldmath${y}$}},{\mbox{%
\boldmath${							\theta}$}}_{0}\right)}{\partial{\mbox{\boldmath${d}$}}}%
Var\left({\ \mbox{\boldmath${\widehat{d}}$}}\right)\left(\frac{\partial%
\widehat{{\ \mbox{\boldmath${T}$}}}\left({\mbox{\boldmath${y}$}},{%
\mbox{\boldmath${							\theta}$}}_{0}\right)}{\partial{\mbox{\boldmath${d}$}%
}}\right)^{\prime}},
\end{equation*}
where,
\begin{equation*}
Var\left(\widehat{{\mbox{\boldmath${d}$}}}\right)=\frac{1}{J}\left[
\Omega_{0}+\sum_{k=1}^{q}\left(1-\frac{k}{q+1}\right)\left(\Omega_{k}+
\Omega_{k}^{\prime}\right)\right],
\end{equation*}
\begin{equation*}
\Omega_{k}=\frac{1}{J}\sum_{g=k+1}^{J}\left({\mbox{\boldmath${d}$}}
^{\left(j\right)}-\widehat{{\mbox{\boldmath${d}$}}}\right)\left({\ %
\mbox{\boldmath${d}$}}^{\left(j\right)}-\widehat{{\mbox{\boldmath${d}$}}}
\right)^{\prime},
\end{equation*}
\begin{equation*}
{\mbox{\boldmath${d}$}}^{\left(j\right)}=\left(d_{3}^{\left(j
\right)},d_{2}^{\left(j\right)}\right)^{\prime}.
\end{equation*}

\item For the Bayes factor, it can be calculated as
\begin{equation*}
\log BF_{10}=\log p\left({\mbox{\boldmath${y}$}}|H_{1}\right)-\log p\left({\ %
\mbox{\boldmath${y}$}}|H_{0}\right).
\end{equation*}
Then following Chib(1995),
\begin{equation*}
\log p\left({\mbox{\boldmath${y}$}}|H_{1}\right)=\log p\left({\ %
\mbox{\boldmath${y}$}}|\bar{{\mbox{\boldmath${\vartheta}$}}}\right)+\log
p\left({\mbox{\boldmath${\bar{\vartheta}}$}}\right)-\log p\left({\ %
\mbox{\boldmath${\bar{\vartheta}}$}}|{\mbox{\boldmath${y}$}}\right),
\end{equation*}
\begin{equation*}
\log p\left({\mbox{\boldmath${y}$}}|H_{0}\right)=\log p\left({\ %
\mbox{\boldmath${y}$}}|{\mbox{\boldmath${\bar{\vartheta}}$}}_{0}\right)+\log
p\left({\mbox{\boldmath${\bar{\vartheta}}$}}_{0}\right)-\log p\left({\ %
\mbox{\boldmath${\bar{\vartheta}}$}}_{0}|{\mbox{\boldmath${y}$}}\right).
\end{equation*}
We can approximate the right-hand side as follows.

\begin{itemize}
\item We use the auxiliary particle filter method proposed by Pitt and
Shephard (1999) to estimate and $\log p\left({\mbox{\boldmath${y}$}}|\bar{{\ %
\mbox{\boldmath${\vartheta}$}}}\right)$ and $\log p\left({\ %
\mbox{\boldmath${y}$}}|\bar{{\mbox{\boldmath${\vartheta}$}}}_{0}\right)$.
The code is provided by Creal (2012).

\item $\log p\left({\mbox{\boldmath${\bar{\vartheta}}$}}\right)$ and m$\log
p\left({\mbox{\boldmath${\bar{\vartheta}}$}}_{0}\right)$ are easy to
evaluate since the prior distributions are standard statistical
distributions.

\item For $\log p\left({\mbox{\boldmath${\bar{\vartheta}}$}}|{\ %
\mbox{\boldmath${y}$}}\right)$ and $\log p\left({\mbox{\boldmath${%
\bar{						\vartheta}}$}}_{0}|{\mbox{\boldmath${y}$}}\right)$, we can use
the approach  of Chib (1995) to estimate them.
\end{itemize}

However, since the NSE of the logarithmic observed likelihood function
dominates that of the logarithmic marginal likelihood which is estimated by
particle filters, the NSE of the BF cannot be obtained.
\end{itemize}

\section*{{\protect\huge \textbf{References}}}

\begin{description}
\item An, S. and Schorfheide, F. (2007). Bayesian Analysis of DSGE Models.
\textit{Econometric Reviews}, \textbf{26}, 211-219.

\item Bao, Y. and Ullah, A. (2007). The second-order bias and mean squared
error of nonlinear estimator in time series. \textit{International
Statistical Review}, \textbf{70},351-372.

\item Bernardo, J.M. and Rueda, R. (2002). Bayesian hypothesis testing: A
reference approach. \textit{International Statistical Review}, \textbf{70},
351-372.

\item Bester, C.A. and Christian, H. (2006). Bias reduction for bayesian and
frequentist estimators. \emph{SSRN Working Paper Series}


\item Bickel,P.J. and Doksum,K. (2006). \textit{Mathematical Statistics:
Basic Concepts and Selected Ideas}, Vol.I,second version. Prentice
Hall,Upper Saddle River,NJ.

\item Berger, J.O., and Perrichi, L.R. (1996). The intrinsic Bayes factor
for Model Selection and Prediction. \textit{Journal of the American
Statistical Association}, \textbf{91}, 109-122.

\item Bernardo, J.M. and Rueda, R. (2002). Bayesian hypothesis testing: A
reference approach. \textit{International Statistical Review}, \textbf{70},
351-372.

\item Black, F.(1976). Studies of stock market volatility changes. \textit{%
Proceedings of the American Statistical Association, Business and Economic
Statistics Section}, 177-181.

\item Geweke, J. Koop,G. and van Dijk, H. (2011). The Oxford Handbook of
Bayesian Econometrics. Oxford University Press.

\item Ghosh, J. and Ramamoorthi, R. (2003). \textit{Bayesian Nonparametrics}%
, Springer Verlag.

\item Han, C. and Carlin, B.P. (2001). Markov chain Monte Carlo methods for
computing Bayes factor: a comparative review. \textit{Journal of the
American Statistical Association}, \textbf{96(455)}, 1122-1132.

\item Imai,S., JAIN, N., and Ching A. (2009). Bayesian Estimation of Dynamic
Discrete Choice Models, \textit{Econometrica}, \textbf{77}, 1865-1899.

\item Kass, R. E. and Raftery, A. E. (1995). Bayes Factors. \textit{Journal
of the Americana Statistical Association}, \textbf{90}, 773-795.

\item Kass, R. E., Tierney, L. and Kadane, J. B. (1990). The validity of
posterior expansions based on Laplace method. \textit{Bayesian and
likelihood methods in statistics and econometrics: Essays in honor of George
A. Barnard}, ed. by Geisser, S., Hodges, J. S., Press, S. J., and Zellner,
A. Elsevier Science Publishers B.V.: North-Holland, \textbf{7}, 473-488.

\item Le Cam,L. and Yang,G.L. (2000). \textit{Asymptotics in Statistics:
Some Basic Concepts,} second edition. Springer-Verlag,New York.

\item Li, Y. and Yu, J. (2012). Bayesian hypothesis testing in latent
variable models. \textit{Journal of Econometrics}, 166(2), 237-246.

\item Li, Y., Zeng, T. and Yu, J. (2014). A new approach to Bayesian
hypothesis testing. \textit{Journal of Econometrics}, \textbf{178(3)},
602-612.

\item Li,Y.,Liu,X.B. and Yu,J. (2015). A Bayesian Chi-Squared Test for
Hypothesis Testing. \textit{Journal of Econometrics}, \textbf{189(1)}, 54-69.

\item Miyata,Y. (2004) Fully exponential Laplace approximations using
asymptotic modes. \textit{Journal of the American Statistic Association},
\textbf{81}, 82-86.

\item Miyata,Y. (2010) Laplace approximations to means and variances with
asymptotic modes. \textit{Journal of statistical planning and inference},
\textbf{140}, 382-392.

\item O'Hagan. (1995). Fractional Bayes Factors for Model Comparison (with
discussion). \textit{Journal of the Royal Statistical Society, Series B}, {%
\textbf{57}, 99-138. }

\item Poirier, D.J. (1995). Intermediate Statistics and Econometrics: A
Comparative Approach. MIT Press, Cambridge, MA.

\item Poirier, D.J. (1997). A predictive motivation for loss function
specification in parametric hypothesis testing. \textit{Economic letters},
\textbf{56}, 1-3.

\item Rilstone, P., Srivatsava,V.K., and Ullah, A. (1996). The second order
bias and MSE of nolinear estimators . \textit{Journal of Econometrics},
\textbf{75}, 369-395.

\item Robert, C. (1993). A note on Jeffreys-Lindley Paradox. \textit{%
Statistica Sinica}, \textbf{3}, 601-608.

\item Tanner, T.A. and Wong, W.H. (1987). The calculation of posterior
distributions by data augmentation. \textit{Journal of the American
Statistical Association}, \textbf{82}, 528-540.

\item Ullah, A. (2004). Finite sample econometrics. Oxford University Press.

\item Yu, J. (2005). On leverage in a stochastic volatility models. \textit{%
Journal of Econometrics}, \textbf{127}, 165-178.
\end{description}

\end{document}